\documentclass[reprint,amsmath,amssymb,prb]{revtex4-2}
\usepackage{graphicx}
\usepackage{dcolumn}
\usepackage{bm}
\usepackage{enumitem}
\usepackage[version=3]{mhchem}
\usepackage{textcomp}
\usepackage{verbatim}
\usepackage{multirow}
\usepackage{nicefrac}
\usepackage[usenames,dvipsnames]{color}
\definecolor{niceblue}{RGB}{31,119,180}
\definecolor{nicered}{RGB}{214,39,40}
\usepackage[colorlinks,breaklinks,bookmarks=true,citecolor=niceblue,linkcolor=nicered,urlcolor=niceblue]{hyperref}
\def\cred{\color{red}}
\def\cbl{\color{blue}}
\newcommand{\bs}{\boldsymbol}
\newcommand{\be}{\begin{equation}}
\newcommand{\ee}{\end{equation}}
\newcommand{\bea}{\begin{eqnarray}}
\newcommand{\eea}{\end{eqnarray}}
\def\vec{\mathbf}
\def\mc{\mathcal}

\def\upa{\uparrow}
\def\dna{\downarrow}

\begin{document}

\title{
{Geometric frustration and Dzyaloshinskii-Moriya interactions\\
in a quantum star lattice hybrid copper sulfate}\\
}

\author{Hajime Ishikawa}
\email{hishikawa@issp.u-tokyo.ac.jp}
\author{Yuto Ishii}
\author{Takeshi Yajima}
\author{Yasuhiro H. Matsuda}
\author{Koichi Kindo}
\affiliation{%
Institute for Solid State Physics, University of Tokyo, Kashiwa, Chiba, 277-8581, Japan 
}%

\author{Yusei Shimizu}
\affiliation{%
Institute for Materials Research, Tohoku University, Oarai, Ibaraki 311-1313, Japan
}%

\author{Ioannis Rousochatzakis}
\affiliation{Department of Physics, Loughborough University, Loughborough LE11 3TU, UK}

\author{Ulrich K. R\"{o}{\ss}ler}
\author{Oleg Janson}
\email{o.janson@ifw-dresden.de}

\affiliation{
Institute for Theoretical Solid State Physics, Leibniz IFW Dresden, 01069 Dresden, Germany
}%

\date{\today}

\begin{abstract}
We study the magnetism of a layered, spin-$\frac12$ organic-inorganic copper sulfate, which is a close realization of the star lattice antiferromagnet, one of the playgrounds of geometric frustration and resonating valence bond physics in two spatial dimensions. Our thermodynamic measurements show no ordering down to 0.1~K and a characteristic field-induced entropic shift, revealing the presence of an infinite number of competing states down to very low energy scales. The response to external magnetic fields shows, in addition, a peculiar anisotropy, reflected in the formation of a 1/3 magnetization plateau (stable up to full saturation around 105\,T) and a paramagnetic, Curie-like susceptibility for one direction of the field (${\bf H}\parallel{\bf c}$), and a completely different response in other field directions. 
Our first-principles density functional theory calculations and exact diagonalizations show that these experimental puzzles are distinctive signatures of a strong interplay between geometric frustration and sizable Dzyaloshinskii-Moriya interactions, and the emergence of a continuous U(1) symmetry at low energy scales.
\end{abstract}

\maketitle

\begin{figure}[tb]
\includegraphics[width=8.6cm]{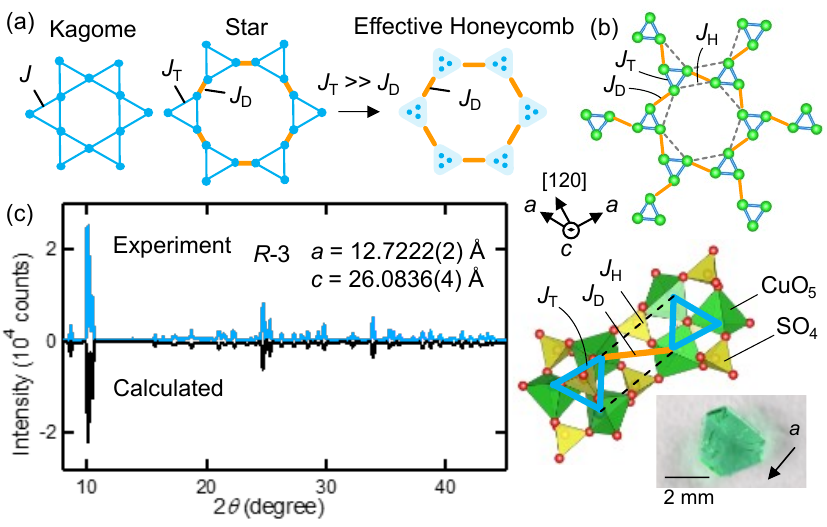}
\caption{\label{fig:star_str_xrd} (a) Kagome and star lattices. Close to the isolated-triangle limit ($J_{\rm{T}}\gg{}J_{\rm{D}}$), the star lattice can be described as honeycomb lattice made of spin triangles. (b) Isotropic exchange  interactions in DiMACuS (top): the $({\bf a},{\bf b},{\bf c})$ frame shown refers to the hexagonal unit cell. Local crystal structure (bottom) described by \ce{CuO5} pyramid and \ce{SO4} tetrahedra, as plotted using VESTA~\cite{momma2011vesta}. A picture of a single crystal is also shown. (c) Observed and calculated powder XRD patterns.} 
\end{figure}

\paragraph*{Introduction.}
The problem of tiling a plane with regular convex polygons fascinated architects since ancient times, engendering a rich cross-cultural heritage. In modern condensed matter physics, the very same patterns appear in a different context: antiferromagnetically coupled localized spins arranged on the vertices of a tiling are at heart of frustrated magnetism~\cite{lacroix2011introduction,richter2004review} and high-temperature superconductivity~\cite{anderson1987resonating}. The emergent spin lattices split into two classes. Bipartite lattices, such as the square or the honeycomb lattice, feature the classical N\'eel ground state (GS), which is globally compatible with all antiferromagnetic (AFM) bonds. Such compatibility is fundamentally impossible in geometrically frustrated lattices, where only local constraints can be satisfied. For instance, in the kagome lattice (Fig.~\ref{fig:star_str_xrd}~a), the local constraint imposes a 120$^{\circ}$-spin structure within a each triangle. Since infinitely many \emph{global} configurations satisfy this \emph{local} constraint, the classical GS manifold is infinitely degenerate. In the case of spin-$\frac12$, quantum fluctuations may stabilize exotic GSs such as quantum spin liquids~\cite{kitaev2006anyons,yan2011spin, depenbrock2012nature} and valence bond phases~\cite{singh2007ground,capponi2013p,Ralko2014}.

The spin-$\frac12$ star lattice antiferromagnet (Fig.~\ref{fig:star_str_xrd}~a) is one of the paradigmatic models of geometrically frustrated magnetism in two spatial dimensions~\cite{richter2004review,lacroix2011introduction,farnell2014quantum}.
As in the kagome, the triangle-based structure leads to an infinite ground state degeneracy at the classical level~\cite{richter2004absence}, and an array of unconventional ground states (GSs) in the spin $S=1/2$ limit, including valence bond solids~\cite{richter2004absence,yang2010competing,ran2018emergent}, resonating valence bond states~\cite{jahromi2018spin}, chiral spin liquids~\cite{yao2007exact}, as well as magnetic field-induced phase transitions~\cite{richter2004absence,ran2018emergent}. 
Unlike the kagome, the star lattice features two inequivalent nearest-neighbour (NN) bonds. The respective Heisenberg exchange interactions are often called $J_{\rm{T}}$ and $J_{\rm{D}}$, as the spins on these bonds are part of a triangle and a dimer, respectively. The lattice can also be viewed as a (decorated) honeycomb lattice made of triangles.

Layered sulfates often provide peculiar spin lattices such as kagome lattice in
natural minerals~\cite{townsend1986triangular,ishikawa2013magnetic} and organic-inorganic materials~\cite{pati2008kagome,powell2009co3}. In 2020, Sorolla \emph{et al.}\ reported \ce{[(CH3)2(NH2)]3Cu3(OH)(SO4)4}$\cdot$\ce{0.24H2O}, (Dimmethylammonium Copper Sulfate, called DiMACuS hereafter), which is the first realization of the spin-$\frac12$ star lattice made of \ce{Cu^2+} (3$\textit{d}^9$) ions~\cite{sorolla2020synthesis}. The crystal structure features inorganic \ce{[Cu3(OH)(SO4)4]^3-} layers separated by dimethylammonium cations and crystal water molecules. In addition to $ J_{\rm{T}} $ and $J_{\rm{D}}$, the next nearest neighbor interaction called $J_{\rm{H}}$ here (Fig.~\ref{fig:star_str_xrd}~b), which form the hexagon, may be present due to the rotation of the triangles. Similarly modified next nearest neighbor interactions are found in a classical kagome AFM and their effect on the GS has been examined previously~\cite{ishikawa2014kagome}. Sorolla \emph{et al.} observed paramagnetic behavior down to 1.8\,K in spite of AFM Weiss temperature of 41\,K and proposed DiMACuS as a quantum spin liquid candidate~\cite{sorolla2020synthesis}.

In this Letter, we report the successful synthesis of millimeter-size single crystals of DiMACuS as well as pure powder suitable for detailed magnetic characterizations. We performed magnetization measurements on single crystals down to 0.1\,K, and measured the entire magnetization process of the powder sample in pulsed magnetic fields up to 120\,T.  We also measured the specific heat with and without magnetic fields. Two remarkable experimental results are the absence of magnetic ordering down to 0.1\,K and a peculiar magnetization anisotropy, with a 1/3 plateau visible only in ${\bf H}\!\parallel\!{\bf c}$. 
By combining first-principles calculations with analytical and numerical model simulations, we show that DiMACuS is a quantum star lattice magnet with $J_{\rm{T}} \gg J_{\rm{D}}$ and substantial chiral Dzyaloshinskii-Moriya (DM) interactions within the spin triangles.

\paragraph*{Synthesis.}
The sample was prepared by reacting 0.4\,g of \ce{CuSO4}$\cdot$\ce{5H2O},
0.5\,ml of sulfuric acid and 5\,ml of N,N-dimethylformamide in a glass vial at
80\textdegree C for a few days. Aggregation of bright green crystals including
hexagonal plates of a few millimeter size is formed (Fig.~\ref{fig:star_str_xrd}~b). 
The aggregation is recovered by decantation and washed by
N,N-dimethylformamide.  The sample is immediately vacuum dried and stored
inside an argon-filled glove box as the crystal is hygroscopic. The crystal is
covered by Apiezon-N grease or sealed inside a plastic tube when handling in
air for measurements. Powder x-ray diffraction measurement is performed on the
crushed crystals by a diffractometer with Cu-K$ \alpha_{1} $
radiation (Smart Lab, Rigaku). The observed pattern matches well with the
calculated pattern of DiMACuS with the space group $ \textit{R}\bar{3} $ and
lattice constants $ \textit{a} $ = 12.7222(2) and $ \textit{c} $ = 26.0836(4)
\AA, indicating the successful synthesis: the pattern calculation is performed
by Fullprof software \cite{rodriguez1993recent} including the effect of
preferred orientation along $ \textit{c} $-axis. 

\begin{figure}[tb]
\includegraphics[width=8.6cm]{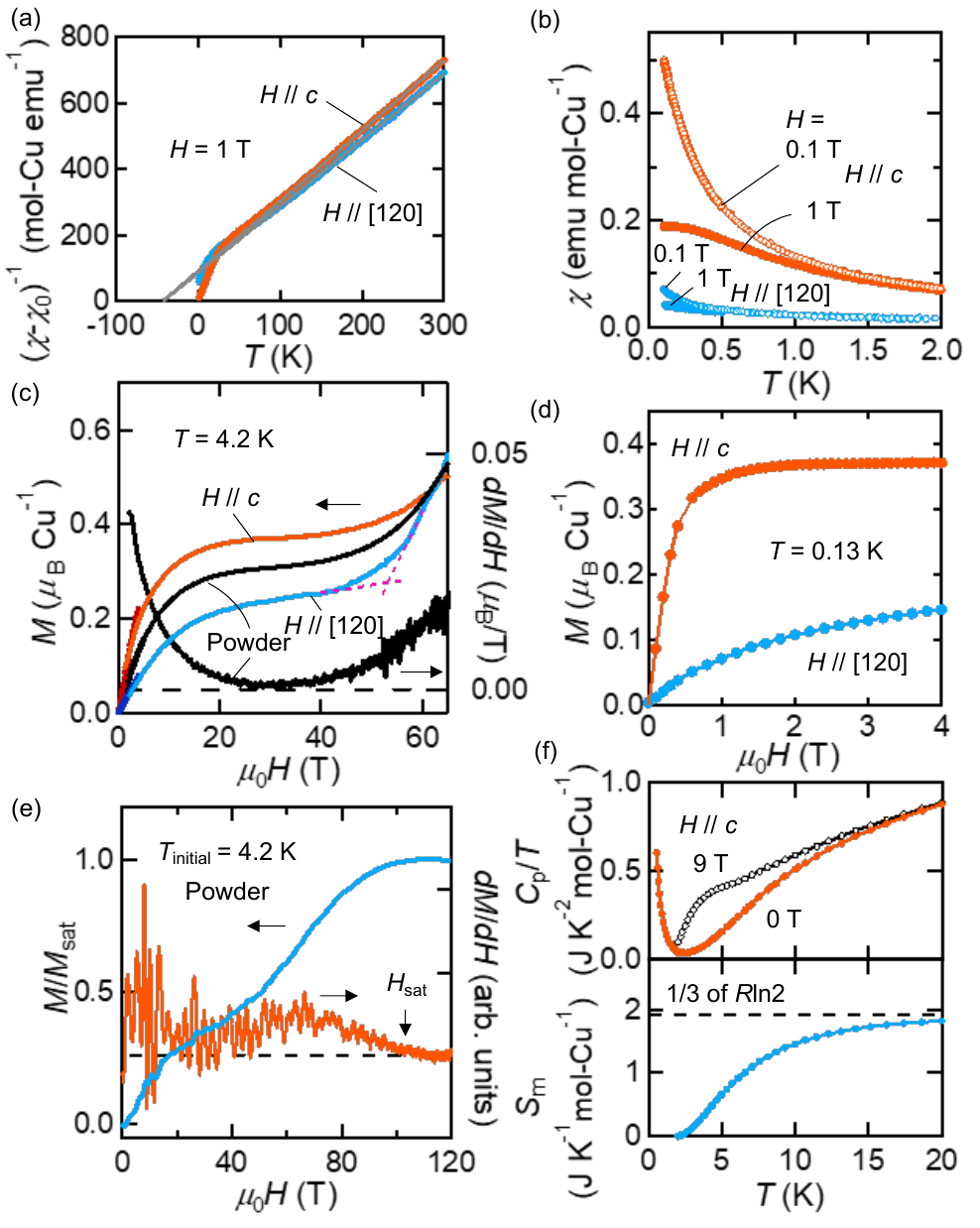}
\caption{\label{fig:chit_mh_cpt} (a) $T$-dependence of inverse susceptibility at 1\,T and the Curie-Weiss fits. (b) Magnetic susceptibility at 0.1 and 1\,T below 2 K. (c) Magnetization curves of the single crystal and powder sample at 4.2 K and up to 65\,T. The magnetic field derivative for powder data is also shown. (d) Magnetization curve of the single crystal at 0.13 K. (e) Magnetization curve of the powder sample up to 120 T measured at $T_{\text{initial}}$ = 4.2\,K and its field derivative. (f) Specific heat divided by $T$ at 0 and 9\,T below 20\,K and magnetic entropy estimated as in the main text.}
\end{figure}

\paragraph*{Low-field magnetization measurements.} 
The magnetization of a single crystal was measured by a SQUID magnetometer (MPMS-XL, Quantum Design) in the $T$-range 1.8--300 K and magnetic fields of up to 5\,T. We made a Curie-Weiss fit to the magnetic susceptibility $\chi$ using the expression $\chi(\textit{T})\!=\!\chi_{0}\!+\!\textit{C}/(\textit{T}\!+\!\Theta)$, where $\textit{C}$ and $\Theta$ are the Curie constant and Weiss temperature, respectively, and $\chi_{0}$ is the $T$-independent term. Above 150\,K, we obtain $C$=0.469(8) emu/mol-Cu$\cdot$K ($\mu_{\rm{eff}}$=1.94 $\mu_{\rm{B}}$), $\Theta$=43(2)\,K, and $\chi_{0}=-3.0(1)\times10^{-4}$ emu/mol-Cu for ${\bf H}\!\parallel\!{\bf c}$. The same fit for the data in ${\bf H}\!\parallel\!{[120]}$, parallel to the edge of the hexagonal crystal, yields a similar result with slightly larger Curie constant: $C$=0.498(8) emu/mol-Cu$\cdot$K ($\mu_{\rm{eff}}$=1.99 $\mu_{\rm{B}}$), $\Theta$=43(2)\,K and $\chi_{0}=-3.0(1)\times10^{-4}$ emu/mol-Cu. Note that no clear anisotropy was observed in $\chi(\textit{T})$ in ${\bf H}\!\parallel\!{[120]}$ and ${\bf H}\!\parallel\!{\bf a}$ above 1.8 K. The linear behavior of the inverse susceptibility $(\chi - \chi_{0})^{-1}$ justifies the fits (Fig.~\ref{fig:chit_mh_cpt}~a). The enhancement of $\mu_{\rm{eff}}$ with respect to the spin-only value (1.73 $ \mu_{\rm{B}}$) is typical for \ce{Cu^2+} compounds and shows a slight anisotropy in the $g$ factor ($g_\parallel\!\simeq\!2.24$ vs $g_\perp\!\simeq\!2.30$).

Below $\sim$50 K, the susceptibility starts to deviate from the Curie-Weiss behavior (Fig.~\ref{fig:chit_mh_cpt}~a). In accord with a previous study~\cite{sorolla2020synthesis}, neither anomalies indicative of magnetic ordering nor signatures of a spin gap formation were observed down to 1.8\,K. To obtain more information on the low-$T$ regime, we measured magnetization of the single crystal samples down to 0.1\,K by the capacitive Faraday method~\cite{shimizu2021development}. Again, the results do not show any sign of magnetic ordering, despite the relatively large $\Theta$ value (Fig.~\ref{fig:chit_mh_cpt}~b). This is one of the key experimental results. Furthermore, the low-$T$ susceptibility is highly anisotropic and is significantly suppressed in ${\bf H}\!\parallel\!{[120]}$ than in ${\bf H}\!\parallel\!{\bf c}$. More importantly, at the lowest field measured (0.1 T), $\chi_\parallel$ exhibits paramagnetic behaviour ($\chi_\parallel\!\propto\!1/T$) down to 0.1\,K, whereas $\chi_\perp$ appears to saturate at low $T$. So the longitudinal and transverse responses are {\it qualitatively} different. This is another key experimental result, which suggests the presence of anisotropic interactions. 

\paragraph*{High-field magnetization measurements.} 
In a magnetic field, the star lattice models can feature exotic behavior with a cascade of phases~\cite{richter2004absence,ran2018emergent}. To study the emergence of field-induced phases in DiMACuS, we performed magnetization measurements in a pulsed high magnetic field. Measurements at 4.2 K up to 65\,T were performed by the induction method in a magnetic field with the pulse length of 4 milliseconds (Fig.~\ref{fig:chit_mh_cpt}~c). As in the magnetic susceptibility, \textit{M} is much larger in ${\bf H}\!\parallel\!{\bf c}$ than in ${\bf H}\!\parallel\!{[120]}$, with the powder data taking intermediate values. While the magnetization increases rapidly below 20\,T, its slope $\frac{\partial{}M}{\partial{}H}$ first decreases as the field increases, but then saturates around 30\,T, which is reminiscent of a magnetization plateau. 

This feature is examined in more detail by measuring the magnetization process at 0.13\,K (Fig.~\ref{fig:chit_mh_cpt}~d, inset). In ${\bf H}\!\parallel\!{\bf c}$, the magnetization  increases steeply and becomes essentially flat above 2.5\,T (with $\frac{\partial{}M}{\partial{}H}$ exhibiting a saturating behaviour already at 1\,T), at around 0.37\,$\mu_{\rm{B}}$/Cu. With $g_\parallel\!\simeq\!2.24$ (estimated from the Curie-Weiss fit) this is close to $\frac13$ of the full moment. Therefore, we conclude that DiMACuS exhibits a $\frac13$ magnetization plateau in ${\bf H}\!\parallel\!{\bf c}$. On the other hand, the magnetization in ${\bf H}\!\parallel\!{[120]}$ gradually increases without a clear flat region. The presence of a clear 1/3 plateau for ${\bf H}\!\parallel\!{\bf c}$ despite the presence of anisotropic interactions 
is the third key experimental result.

The end of the plateau-like regime is signalled by the increase of $\frac{\partial{}M}{\partial{}H}$ which becomes noticeable around 40\,T at 4.2\,K. A more evident increase of magnetization is observed in ${\bf H}\!\parallel\![120]$, where magnetization is suppressed at low magnetic fields. The extrapolated (dashed) lines (Fig.~\ref{fig:chit_mh_cpt}~c) cross at 53~T, which may be considered as the onset of the jump to saturation for ${\bf H}\!\perp\!{\bf c}$. To study the behavior in higher fields, we performed magnetization measurements by the induction method up to around 120\,T (Fig.~\ref{fig:chit_mh_cpt}~e) generated by the destructive single turn coil method with the pulse length of 7 microseconds~\cite{takeyama2011precise}. Powder sample was used due to the small sample space. $\frac{\partial{}M}{\partial{}H}$ exhibits a minimum at around 30\,T and a maximum at around 65 T, which are consistent with the data obtained in the measurements up to 65\,T, albeit with large noise caused by the magnetic field generation. The field derivative becomes small and almost constant above 105\,T, indicating that the fully polarized state is reached.  Indeed, the magnetization at 105\,T is approximately three times larger than the value at 30\,T in the $\frac13$ plateau-like region.

\paragraph*{Specific heat measurements.} 
The specific heat $C_p$ of the single crystal was measured by the relaxation method using a commercial apparatus (PPMS, Quantum Design). At zero field, $C_p/T$ shows a low-$T$ upturn (Fig.~\ref{fig:chit_mh_cpt}~f), which may indicate long-range ordering below 0.1 K (the lowest $T$ in $\chi$ measurements). Measurements in 9 T applied along the \textit{c}-axis reveal a considerable enhancement of $C_p/T$ above 2 K. While we do not have data below 2 K at 9 T, the system is already in the $\frac{1}{3}$ plateau phase (Fig.~\ref{fig:chit_mh_cpt}\,d), and the remaining magnetic entropy should be small. Hence, the field-induced enhancement of $C_p/T$  above 2 K gives strong evidence for the existence of low-lying magnetic excitations residing below 2 K in zero field and which are propelled to higher energies by the field. This behaviour is a hallmark of isolated or weakly-interacting spin-$S$ degrees of freedom emerging at low-energy scales. To shed further light on this, we estimate the entropy content of the low-lying excitations by integrating the difference of $C_p/T$ between 9 and 0 T from 2 K up to 20 K. The obtained entropy is approximately 1.8 J/K mol-Cu$^{-1}$ at 20 K, very close to $\frac{1}{3}\textit{R}\ln2$. This value is associated with {\it one} doublet degree of freedom per Cu triangle. This is the fourth key experimental finding.

\paragraph*{Microscopic  modeling.} 
We now set out to develop a microscopic description that  accounts for all experimental findings. Based on the crystal structure of DiMACuS, we can identify three inequivalent exchange paths between neighbouring spins, $J_{\rm{T}}$, $J_{\rm{D}}$, and $J_{\rm{H}}$ (Fig.~\ref{fig:star_str_xrd}~b), which can be estimated by first-principles density-functional-theory (DFT) calculations. To this end, we employed the generalized gradient approximation (GGA)~\cite{perdew1996generalized} as implemented in the full-potential code FPLO version 21~\cite{koepernik1999full}. Following~\cite{sorolla2020synthesis}, we keep only \ce{Cu3(OH)(SO4)4} magnetic layers and discard the organic cations. A uniform background charge was used to retain electroneutrality. Since the experimental structure features the unusually short hydroxyl bond length of about 0.82 \AA, we calculated the optimal hydrogen position ($z/c$ = 0.370053) with respect to the GGA total energy. Next, we performed magnetic supercell calculations using the GGA+$U$ functional with the Coulomb repulsion $U_d$ of $8.5\pm1$\,eV and Hund's exchange $J_d$ of 1\,eV, respectively, and the fully localized limit as the double counting correction. Magnetic exchange integrals were estimated by mapping the GGA+$U$ energies of eight different magnetic configurations onto a classical Heisenberg model; the resulting redundant linear problem was solved by the least-squares method. For $U_d\!=\!9.5$\,eV, we found $J_{\rm{T}}\!=\!81.5$\,K, $J_{\rm{D}}\!=\!5.4$\,K, and $J_{\rm{H}}\!=\!0.3$\,K (practically negligible)~\cite{suppl}. This choice of $U_d$ is justified by the excellent agreement between the calculated Weiss temperature $\Theta\!=\!\frac12J_{\rm{T}}\!+\!\frac14J_{\rm{D}}\!+\!\frac12J_{\rm{H}}\!=\!42.2$\,K and its experimental value.

The $J_{\rm{T}}$--$J_{\rm{D}}$ model in the strong-coupling limit $J_{\rm{T}}$\,$\gg$\,$J_{\rm{D}}$ suggests an effective description in terms of isolated $S=1/2$ Heisenberg triangles. Indeed, the isolated triangle model with $J_{\rm{T}}$\,=\,58.5 K and $g$\,=\,2.205 accounts for the experimental $\chi(T)$ measured in ${\bf H}\!\parallel\!{\bf c}$~\cite{suppl} and reproduces the $\frac13$ plateau, which corresponds to each triangle being in the $S_z\!=\!\frac{1}{2}$ member of the Zeeman-split doublet (a similar situation is observed in the frustrated cuprate volborthite featuring magnetic trimers~\cite{ishikawa2015one,janson2016magnetic}).

\begin{figure}[!t]
    \includegraphics[width=8.6cm]{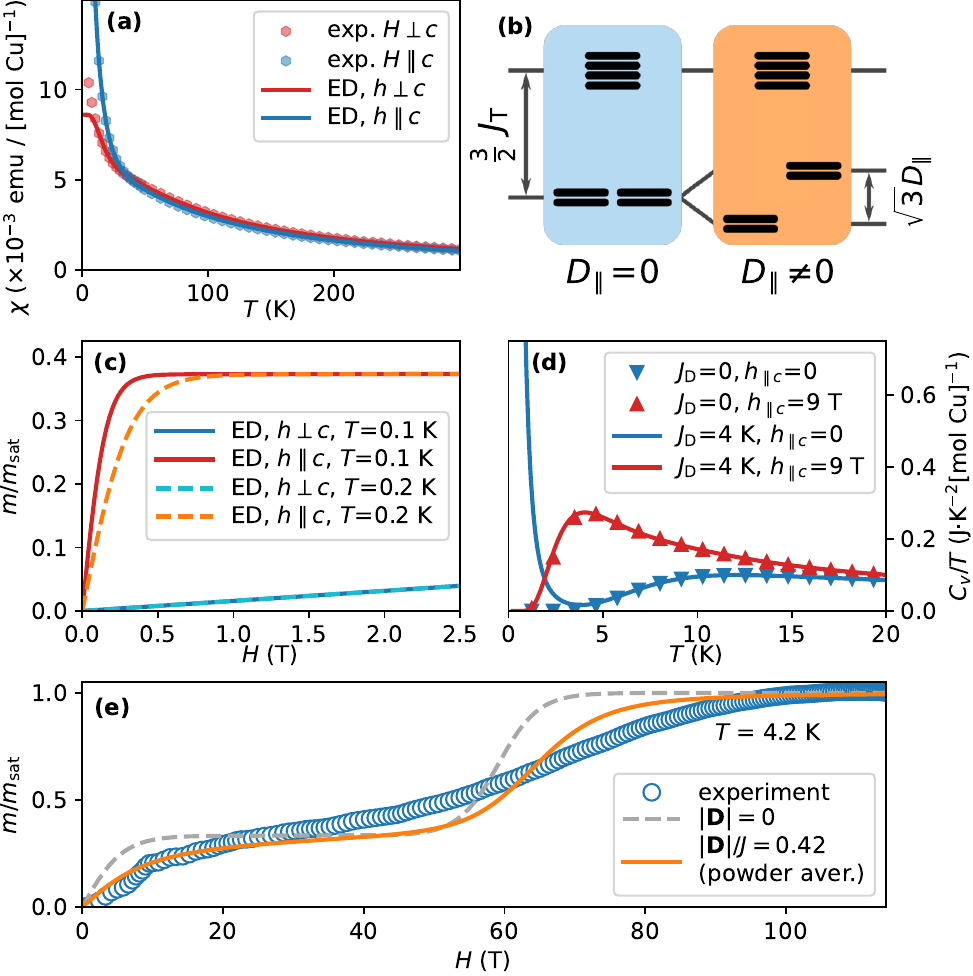}
    \caption{\label{fig:simul} (a) Magnetic susceptibility at 1\,T for different field directions (symbols) and fits (lines) with the anisotropic triangle Heisenberg model with $|{\bf D}|/J\!=\!0.42$ ($J\!=\!56.5$\,K, $g\!=\!2.249$, and $\chi_0\!=\!-2.82\times10^{-4}$ emu\,/\,mol Cu for ${\bf H}\!\perp\!{\bf c}$ and $J\!=\!58.5$\,K, $g\!=\!2.205$, and $\chi_0\!=\!-3.18\times10^{-4}$ emu\,/\,mol Cu for ${\bf H}\!\parallel\!{\bf c}$). (b) Spectra of a Heisenberg triangle and a triangle with $D_{\rm{T}}^\parallel$. (c) Magnetization curve simulated for the parameters from (a).  (d) Simulated specific heat for an isolated triangle ($J_D\!=\!0$) at 0 and 9 T, and for two triangles coupled with $J_D\!=\!4$\,K. (e) Powder-averaged magnetization isotherm of an isolated triangle with $|{\bf D}|/J\!=\!0.42$ (solid line, obtained by averaging over a Fibonacci sphere of 500 points), in comparison with the isotropic triangle (dashed lines) and the experiment (circles). Matplotlib~\cite{matplotlib} was used for plotting (a) and (c)-(e).
    For plotting scripts and numerical data, see Ref.~\cite{repo}.} 
\end{figure}

However, the isotropic model fails to describe the observed anisotropy. Moreover, an isolated $S\!=\!1/2$ AFM Heisenberg triangle has two doublet GSs and not one~\cite{suppl}, and therefore does not capture the $\frac{1}{3}\textit{R}\ln2$ entropy content of the low-lying excitations deduced from the $C_p$ data. To determine the relevant anisotropies, we performed full relativistic noncollinear DFT+$U$ total energy calculations and compute the bilinear exchange matrix $J^{\alpha\beta}_{\rm{T}}$ using the energy mapping method~\cite{xiang13}. Total energies were calculated with the projector-augmented wave code VASP version 5.4.4~\cite{VASP, *VASP_2} using standard pseudopotentials~\cite{VASP_PAW} and the energy cutoff of 400\,eV.  These calculations were done on a 2$\times$2$\times$2 $k$-mesh; for the interaction parameters, we used $U_d$ and $J_d$ of 9.5 and 1\,eV, respectively. In this way, we found that~\cite{suppl}: i) the symmetric and traceless part of the exchange anisotropy is extremely weak and can be safely disregarded, and ii) the dominant Heisenberg exchange in DiMACuS is accompanied by a sizable DM anisotropy on the $J_{\rm{T}}$ bonds (the DM anisotropy on the $J_{\rm{D}}$ bonds vanishes due to inversion symmetry). The ${\bf D}_{\rm{T}}$ vectors are perpendicular to the respective bonds and form acute angles of $\sim$50$^{\circ}$ to each other, with $D_{\rm{T}}^\parallel/J_{\rm{T}}\!=\!0.274$, $D_{\rm{T}}^\perp/J_{\rm{T}}\!=\!0.493$, and $|{\bf D}_{\rm{T}}|/J_{\rm{T}}\!=\!0.56$~\cite{suppl}. The latter ratio is remarkably large, yet not unprecedented for cuprates~\cite{janson2014,panther2023frustration}. The sizable DM is supported by the simulated susceptibility showing excellent agreement for both field directions down to $\sim\!10$\,K (Fig.~\ref{fig:simul}\,a). It allows us to further refine the magnetic exchanges: $J_{\rm{T}}\!=\!57$\,K and $|{\bf D}_{\rm{T}}|/J_{\rm{T}}\!=\!0.42$. 

The DM interactions have a drastic impact on the physics of weakly coupled triangles. The leading contribution stems from $D_{\rm{T}}^\parallel$, while the influence of in-plane DM components is zero to first order in $D_{\rm{T}}^\perp/J_{\rm{T}}$~\cite{suppl}. Our numerical simulations (Fig.~\ref{fig:simul} and \cite{suppl}), which include all DM components, confirm this explicitly. So, to a good approximation, $D_{\rm{T}}^\perp$ can be safely disregarded, and the system effectively features two emergent (i.e., approximate) symmetries at low energies: U(1) spin rotation and a threefold spatial rotation, both around the ${\bf c}$ axis. 

Hence, for ${\bf H}\parallel{\bf c}$, the states of the system have well defined total moment $S_z$ along ${\bf c}$ and chirality $\ell$ (irreps of threefold spatial rotation). Now, in the absence of $D_{\rm{T}}^\parallel$, each triangle has two doublet GSs, separated by the $S\!=\!3/2$ quartet by a gap of $3J_T/2\!\simeq\!85$\,K (Fig.~\ref{fig:simul}\,b, left). At lower temperatures, the $J_T$ scale disappears from the problem, and we are left with two doublets of opposite chirality $\ell$ (irreps of threefold rotation symmetry).  $D_{\rm{T}}^\parallel$ lifts the degeneracy of the two doublets and introduces a new energy scale, the gap $\Delta\!=\!\sqrt{3}|D_{\rm{T}}^\parallel|\!\simeq\!35$\,K, (Fig.~\ref{fig:simul}\,b, right). At lower $T$, this scale also disappears from the problem, and we are left with one doublet per triangle, in agreement with the $C_p$ data (see Fig.~\ref{fig:simul}\,d, which also demonstrates the characteristic field-induced entropic shift seen experimentally, at the level of two coupled triangles).  

The peculiar anisotropy of DiMACuS is governed by the emergent U(1) symmetry: unaffected by a field along ${\bf c}$, but broken for fields perpendicular to ${\bf c}$. This is manifest as follows~\cite{suppl}. $D_{\rm{T}}^\parallel$ gives rise to a characteristic locking between $S_z$ and $\ell$: the $S_z\!=\!\pm1/2$ members of one doublet have $\ell\!=\!\pm1$, whereas for the other doublet $\ell\!=\!\mp1$. Hence, a field along ${\bf c}$ does not couple the two doublets; instead, it gives rise to a Zeeman splitting for each doublet, the Curie-like susceptibility 
$\chi_\parallel\!\propto\!1/(4T)$ at temperatures sufficiently below $\Delta$, and the flat 1/3 magnetization plateau (Fig.~\ref{fig:simul}\,c), independent of $D_{\rm{T}}^\parallel$. 
By contrast, an in-plane field breaks the U(1) symmetry explicitly and thus couples the two doublets, leading to a standard level repulsion, with each doublet retaining its twofold degeneracy. At low-$T$, $\chi_\perp$ approaches a finite value (Fig.~\ref{fig:simul}\,a), inversely proportional to $D_{\rm{T}}^\parallel$,  
and the magnetization behaves as  
$h/\sqrt{h^2+3(D_{\rm{T}}^\parallel)^2}$ (where $h\!=\!g_\perp\mu_BH$): linear at low fields (Fig.~\ref{fig:simul}\,c), and asymptotically approaching the 1/3 value, without developing a flat plateau. Precisely this behavior is observed experimentally in Fig.~\ref{fig:chit_mh_cpt} (c,d).

Next, we consider the high-field magnetization of the isolated-triangle model. On exiting the 1/3 plateau, the magnetization shows an almost abrupt jump to full saturation at $H^\ast$ equal to  $3/2J_{\rm{T}}$ in the 
isotropic case. $D_{\rm{T}}$
propels the saturation to higher fields, but only for ${\bf
H}\!\parallel\!{\bf c}$. 
By using the above refined values of $J_{\rm{T}}$ and $D_{\rm{T}}$ in analytical expressions for $H^\ast$~\cite{suppl}, we obtain $H_\parallel^\ast\!\simeq\!64$\,T and $H_\perp^\ast\!\simeq\!56$\,T. The latter is in satisfactory agreement with the endpoint of the plateau-like region for ${\bf H}\!\perp\!{\bf c}$ (Fig.~\ref{fig:chit_mh_cpt}~c). The difference between $H_\parallel^\ast$ and $H_\perp^\ast$ is further corroborated by numerical simulations of the powder-averaged magnetization (Fig.~\ref{fig:simul}, e). We note finally that, in the powder measurements, the value of $H^\ast$ is distributed between $H_\parallel^\ast$ and $H_\perp^\ast$, effectively destroying the magnetization jumps predicted for both field directions.

The isolated triangle model does not fully account for the specific heat data: the expected gapped zero-field spectrum (Fig.~\ref{fig:simul}\,d) is in sharp contrast with the divergence observed below 0.5\,K (Fig.~\ref{fig:chit_mh_cpt}\,f).  The root cause of this discrepancy are interactions between the triangles: for a minimal model of two triangles coupled with $J_{\rm{D}}$, such a peak readily appears at a temperature very close to the experimentally observed (Fig.~\ref{fig:simul}\,d). Obviously, the actual connectivity of triangles in DiMACuS follows a honeycomb lattice, and its realistic simulation requires a much larger number of spins. The interactions between the triangles can be described by an effective model, in which each spin triangle is treated as a rigid entity.  While a full analysis of this model is beyond our scope, we note that the rescaling of effective spin lengths also reduces the dominant exchange scale from $J_{\rm{D}}$ to $J_{\rm{D}}/9$, which may explain the lack of ordering down to 0.1\,K.

Finally, we comment on the disorder in the dimethylammonium molecule and defect in the crystal water intercalated between the inorganic layers. A slight orientational disorder of the dimethylammonium molecule is reported~\cite{sorolla2020synthesis}. On the organic molecular Mott insulator, there is an argument that random freezing of the electric polarization of the molecules causes magnetic bond randomness, that can result in a gapless spin liquid behavior~\cite{watanabe2014quantum}. In contrast to the organic Mott insulator where the molecule itself carries the spin, the dimethylammonium and \ce{H2O} molecules in DiMACuS are non-magnetic and not involved in the magnetism in the \ce{Cu3(OH)(SO4)4} layer. Indeed, no structural disorder is reported in the \ce{Cu3(OH)(SO4)4} layer~\cite{sorolla2020synthesis}. While we anticipate the effects of disorder on the magnetism to be weak, this may still hinder the ordering at very low $T$.

\paragraph*{Summary and outlook.}
We demonstrate DiMACuS as a realization of a spin-$\frac12$ star lattice antiferromagnet, one of the paradigms for geometric frustration and resonating valence bond physics. The main experimental puzzles, including the absence of magnetic ordering down to very low temperatures, the characteristic field-induced entropic shift seen in the specific heat data, and the peculiar anisotropy in the magnetic response, can all be accounted for by the strong frustration in the Cu triangles and the sizable Dzyaloshinskii-Moriya anisotropy. Further studies are needed to clarify if DiMACuS orders magnetically at a very low temperature and investigate the possible role of disorder.

Star lattice magnets are known to exhibit chiral spin liquid state in the presence of Kitaev-type anisotropic interaction~\cite{yao2007exact}.  Substituting Cu with divalent Co may substantially enhance the exchange anisotropy and give rise to bond-dependent interactions that underlie the Kitaev physics. As demonstrated in a kagome system, cobaltate analogs of cuprates exist in nature~\cite{kasatkin2015karpenkoite} and can be
synthesized~\cite{haraguchi2022perfect}. Synthesis of relative materials with different magnetic cations will pave the way to explore the star lattice magnetism from the extreme quantum case to the classical large-$S$ limit.

\begin{acknowledgments}
We thank Dr.\ D.\ Nishio-Hamane for taking the optical image of the crystal. We thank Prof.\ Z.\ Hiroi for fruitful discussions and U.\ Nitzsche for technical assistance. HI was supported by JSPS KAKENHI Grants No.\ JP22H04467 and No.\ JP 22K13996. OJ was supported by the Leibniz Association through the Leibniz Competition and the German Forschungsgemeinschaft (DFG, German Research Foundation) through SFB 1143 (project-id 247310070). IR acknowledges the support by the Engineering and Physical Sciences Research Council, Grant No.\ EP/V038281/1. 
\end{acknowledgments}

\def\theequation{\arabic{equation}}
\def\thefigure{S\arabic{figure}}
\def\thetable{S\arabic{table}}
\setcounter{equation}{0}
\setcounter{figure}{0}
\setcounter{table}{0}

\bigskip
\begin{widetext}
\newpage
\begin{center}
{
\smallskip
\large
Supplemental Material: \\ \textbf{Geometric frustration and Dzyaloshinskii-Moriya interactions \\ in a quantum star lattice hybrid copper sulfate}

\bigskip

\normalsize
Hajime Ishikawa, Yuto Ishii, Takeshi Yajima, Yasuhiro H. Matsuda, Koichi Kindo, Yusei Shimizu, Ioannis~Rousochatzakis, Ulrich K. R\"{o}{\ss}ler, and Oleg Janson}
\end{center}
\bigskip

In this supplementary material we provide technical derivations and auxiliary information for: i) the first-principles calculations of the isotropic (Heisenberg) and anisotropic, in particular Dzyaloshinskii–Moriya (DM), interactions in DiMACuS (Sec.~\ref{sec:j}), ii) the quantum-mechanical problem of a single spin-1/2 triangle with antiferromagnetic (AF) exchange $J$ and DM interactions (Sec.~\ref{sec:triangle}), and iii) simulations of the magnetic susceptibility of this model and fits to the experimental magnetic susceptibility of DiMACuS that facilitate an accurate estimation of the exchange parameters (Sec.~\ref{sec:ChiFit}).

\section{\label{sec:j}Estimation of the exchange integrals}

\subsection{Isotropic (Heisenberg) exchanges}

In DFT calculations, we consider the experimental crystal structure of
DiMACuS~\cite{sm:sorolla2020synthesis}, remove the organic cation, add a uniform
background charge to ensure the electroneutrality, and use the GGA-optimized
position for H atoms. All calculations are done using the rhombohedral unit
cell, whose volume is three times smaller than that of the hexagonal cell
depicted in Fig.~\ref{fig:str} (left). Estimates for the three leading
exchanges as a function of the Coulomb repulsion $U_d$ are provided in
Table~\ref{tab:exchanges}. The value of $U_d=9.5$\,eV gives an excellent agreement with
the experimental Weiss temperature.

\begin{figure}[htb]
\centering
\includegraphics[width=.8\textwidth]{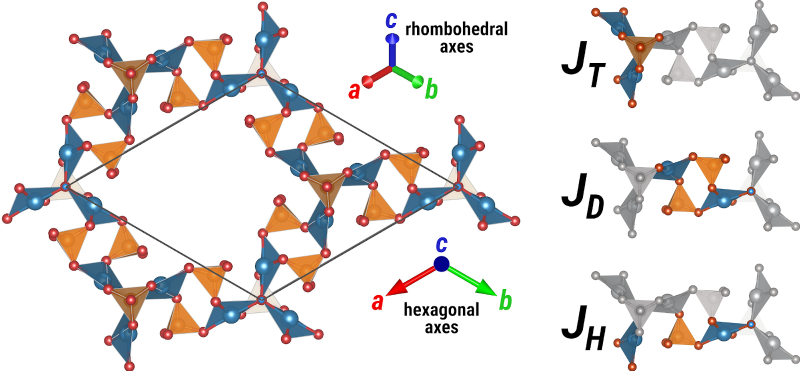}
\caption{\label{fig:str}
Left: Crystal structure of the magnetic Cu$_3$(OH)(SO$_4$)$_4$ layers in DiMACuS, featuring distorted CuO$_4$ plaquettes (blue) and SO$_4$ tetrahedra (orange). The unit cell parameters are $a=b=12.7222$\,\r{A}, $c=26.0836$\,\r{A}, $\alpha=\beta=90^{\circ}$, $\gamma=120^{\circ}$ (hexagonal axes) or $a=b=c=11.31263$\,\r{A}, $\alpha=\beta=\gamma=68.435^{\circ}$ (rhombohedral axes). All calculations are done for the rhombohedral cell. The hexagonal unit cell is plotted with gray lines. 
Right: Crystalline environments facilitating the three leading in-plane exchange couplings $J_{\rm{T}}$, $J_{\rm{D}}$, and $J_{\rm{H}}$. The crystal structure was plotted using vesta~\cite{sm:momma2011vesta}.}
\end{figure}

\begin{table}[h]
    \caption{\label{tab:exchanges}
    Exchange integrals $J_{\rm{T}}$, $J_{\rm{D}}$, $J_{\rm{H}}$ and the Weiss
temperature $\theta$ (all in K) estimated from GGA+$U$ ($U_d$\,=\,9.5\,eV, $J_d$\,=\,1\,eV) total energies for 8 magnetic configurations.}
    \begin{ruledtabular}
	\begin{tabular}{rdddddd}
	    \multirow{3}{*}{exchange} & \multirow{3}{*}{bonds per spin} & \multirow{3}{*}{$d_{\text{Cu..Cu}}$ (\r{A})} & \multicolumn{4}{c}{$J$ (K)} \\
            & & & \multicolumn{3}{c}{FPLO} & \multicolumn{1}{c}{vasp} \\
            & & & U_d=7.5 & U_d=8.5 & U_d=9.5 & U_d=9.5 \\ \hline
	    $J_{\rm{T}}$ &          1  & 3.2797 &   124.5 &   101.3 &     \bf{81}.\bf{5} & 77.1 \\
	    $J_{\rm{D}}$ & \nicefrac12 & 4.4387 &     8.5 &     6.8 &     \bf{5}.\bf{4}  &  4.1 \\
	    $J_{\rm{H}}$ &          1  & 6.0010 &     0.2 &     0.3 &     \bf{0}.\bf{3}  &  0.4 \\ \hline
	    $\theta$     &             &        &    64.5 &    52.5 &    \bf{42}.\bf{2}  &  39.8\\
        \end{tabular}
    \end{ruledtabular}
\end{table}

\subsection{Anisotropic components of $J_{\rm{T}}$}
The experimentally observed anisotropy of the magnetic susceptibility despite the similarity of the Weiss temperature for both field directions hints at possible relevance of the antisymmetric DM exchange $\vec{D}_{\rm{T}}$. To estimate this exchange numerically, we determine all elements of the respective bilinear exchange matrix $\mathbf{M}_{\rm{T}}$ by using the energy mapping method~\cite{sm:xiang13}, and then decomposing this matrix as
\be
{\bf M}_{\rm{T}} = 
J_{\rm{T}}~{\bf 1} + \bs{\Sigma}_{\rm{T}} + {\bf A}_{\rm{T}}\,,
\ee
where ${\bf 1}$ is the 3$\times$3 identity matrix and
\be
J_{\rm{T}}=\frac{1}{3}~\text{Tr}[\mathbf{M}_{\rm{T}}]
\ee
is the Heisenberg coupling, 
\be
\bs{\Sigma}_{\rm{T}}=\frac{1}{2}\left(
{\bf M}_{\rm{T}}+{\bf M}_{\rm{T}}^{\top}
\right)
-\frac{1}{3}~\text{Tr}[{\bf M}_{\rm{T}}]~{\bf 1}
=\bs{\Sigma}_{\rm{T}}^{\top}
\ee
is the symmetric and traceless part of ${\bf M}_{\rm{T}}$ (${\bf M}_{\rm{T}}^{\top}$ is the transpose matrix of ${\bf M}_{\rm{T}}$), and 
\be
{\bf A}_{\rm{T}}=\frac{1}{2}\left({\bf M}_{\rm{T}}-{\bf M}_{\rm{T}}^{\top}\right)=-{\bf A}_{\rm{T}}^{\top}
\ee
is the anti-symmetric part of ${\bf M}_{\rm{T}}$, which is related to the components of ${\bf D}_{\rm{T}}$ as $A^{\alpha\beta}_{\rm{T}} = \epsilon^{\alpha\beta\gamma} D_{\rm{T}}^\gamma$, or, more explicitly, 
\be
{\bf A}_{\rm{T}}= 
\begin{pmatrix}
0&D_{\rm{T}}^Z&-D_{\rm{T}}^Y\\
-D_{\rm{T}}^Z&0&D_{\rm{T}}^X\\
D_{\rm{T}}^Y&-D_{\rm{T}}^X&0\end{pmatrix}\,,
\ee
where $X$, $Y$, $Z$ refer to the crystallographic (Cartesian) axes.

\begin{table}[tb]
\caption{\label{tab:cu_sites}Positions of selected atoms in the unit cell in the crystallographic (Cartesian) frame (Fig.~\ref{fig:dm_vecs}).}
\begin{ruledtabular}
\begin{tabular}{rdddp{.05\textwidth}rddd}
site & \multicolumn{1}{c}{$X$ (\AA)} &  \multicolumn{1}{c}{$Y$ (\AA)} & \multicolumn{1}{c}{$Z$ (\AA)} & & site & \multicolumn{1}{c}{$X$ (\AA)} &  \multicolumn{1}{c}{$Y$ (\AA)} & \multicolumn{1}{c}{$Z$ (\AA)}\\ \hline
Cu(1) & 1.571222 & 3.828742 & 7.097030 & & Cu(3) & 4.579598 & 2.909116 & 6.169293\\
Cu(2) & 3.265568 & 5.665752 & 4.973043 & & S(7)  & 2.196772 & 2.893669 & 4.255106
\end{tabular}
\end{ruledtabular}
\end{table}

\begin{figure}[!b]
\centering
\includegraphics[width=.45\textwidth]{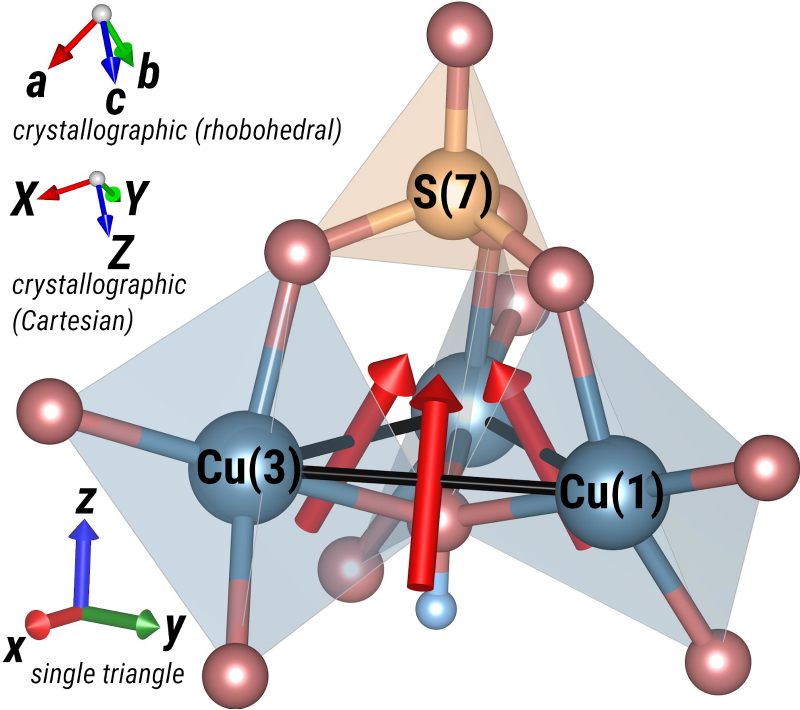}
\caption{\label{fig:dm_vecs}
Local structure of the Cu(1)-Cu(2)-Cu(3) triangles, with the S(7)O4
tetrahedron (shaded orange) above the plane of the spins. The red arrows show
the orientations of the DM vectors on the nearest-neighbour bonds around the
triangle. We also show three different frames: i) The crystallographic $(a, b,
c)$ frame corresponding to the rhombohedral unit cell. This frame differs from
the $({\bf a},{\bf b}, {\bf c})$ frame of the main text, which corresponds to
the hexagonal unit cell. ii) The Cartesian $(X,Y,Z)$ frame pertains to the
crystallographic coordinate system (same as in Table~\ref{tab:cu_sites}). iii)
The coordinate system $(x,y,z)$ of a single triangle, with the S(7)O4
tetrahedron sitting above the plane of the spins. The transformation $(X,Y,Z)
\rightarrow (x,y,z)$ is described by $R_{\Delta}$ in Eq.~(\ref{eq:r_delta}). The structural elements were plotted using vesta~\cite{sm:momma2011vesta}.}
\end{figure}

The crystal structure of DiMACuS imposes no symmetry constraints on the DM vector $\vec{D}_{\rm{T}}$, but all vectors on different bonds in the cell are connected by symmetry elements. In particular, the three vectors belonging to the same triangle are connected by $C_3$ symmetry. In the triangle formed by the atoms Cu(1), Cu(2), and Cu(3) (Table~\ref{tab:cu_sites}), we consider the first interatomic vector $\vec{r}_{\text{Cu(1)-\text{Cu(3)}}}$. After performing 36 full-relativistic DFT+$U$ total energy calculations, we obtain the following bilinear exchange matrix:
\be
\mathbf{M}_{\rm{T}}^{\text{Cu(1)-\text{Cu(3)}}} =
\begin{pmatrix}
  66.0  & -35.9 & \phantom{+0}3.1\\
  35.9  &  \phantom{+}65.5 & -11.3\\
  -1.6  &  \phantom{+}11.6 & \phantom{+}69.1
\end{pmatrix} \text{K}\,,
\ee
from which we get

\be
J_{\rm{T}}=66.9~\text{K}\,,
\ee

\be\label{eq:d13_cryst}
\vec{D}_{\rm{T}}^{\text{Cu(1)-\text{Cu(3)}}} = \left(-11.5, -2.3, -35.9\right) \text{K}\,,
\ee
and
\be
\bs{\Sigma}_{\rm{T}}^{\text{Cu(1)-\text{Cu(3)}}} =
\begin{pmatrix}
  -0.9  & 0 & 0.8\\
  0  &  -1.3 & 0.2\\
  0.8  &  0.2 & 2.2
\end{pmatrix} \text{K}\,,
\ee
in the $(X, Y, Z)$ frame. So the elements of the symmetric and traceless part of the exchange anisotropy are extremely weak compared to the antisymmetric part, and can therefore be disregarded.

In the following, it is more convenient to resort to a different coordinate system, denoted by $(x, y, z)$ in Fig.~\ref{fig:dm_vecs}, with the midpoint of Cu(1)-Cu(2)-Cu(3) triangle at the origin and the $z$ axis coinciding with the $C_3$ axis which crosses the midpoint and the neighboring S(7) atom. Importantly, there are two such frames, one for the Cu triangles with the S(7) atom above the Cu plane, and another for the Cu triangles with S(7) atoms below the Cu plane. The two frames are related by the inversion center in the middle of the $J_D$ bond that connects the two types of triangles. In the following, we shall use the $(x,y,z)$ frame of the Cu triangles with S(7) above the Cu plane as our global frame. The directions of the DM vectors on the Cu triangles with S(7) atoms below the Cu plane can be obtained by applying inversion symmetry (see site-labelling convention indicated by red arrows in Fig.~\ref{fig:EffCouplings2}).

\begin{figure}[h]
\includegraphics[width=0.45\linewidth]{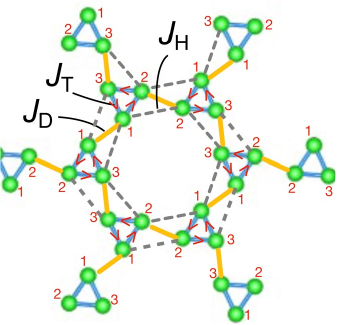}
\caption{Lattice structure of Cu atoms in DiMACus, showing the most relevant Heisenberg couplings. Red arrows indicate our site-labelling convention $i\!\to\!j$ associated with the DM coupling term ${\bf D}_{ij}\cdot{\bf S}_i\times{\bf S}_j$ between spin sites $i$ and $j$. For the actual directions of these vectors see Fig.~\ref{fig:dm_vecs} and Eqs.~(\ref{eq:Dvectors1})-(\ref{eq:Dvectors3}).}\label{fig:EffCouplings2}
\end{figure}

The crystallographic (Cartesian) coordinates are transformed into this new coordinate system
by the following matrix:
\be\label{eq:r_delta}
\mathbf{R}_{\Delta} = \begin{pmatrix}
\phantom{+}0.35633105& -0.84786434&  \phantom{+}0.39262481\\
-0.84786434& -0.11683799&  \phantom{+}0.51717980\\
-0.39262481& -0.51717980&  -0.76050695\end{pmatrix}\,.
\ee
By shifting the origin and applying this rotation matrix to the atomic coordinates from Table~\ref{tab:cu_sites}, we obtain the coordinates in the $(x,y,z)$ frame:
\begin{equation}
\begin{split}
\vec{r}_1 &= \left(0.1001, 1.8909, 0\right) \text{\AA}, \\
\vec{r}_2 &= \mathbf{R}_z(\nicefrac{2\pi}{3})\cdot\vec{r}_1 = \left(-1.6876, -0.8588, 0\right)\text{\AA}, \\
\vec{r}_3 &= \mathbf{R}_z(-\nicefrac{2\pi}{3})\cdot\vec{r}_1 =  \left(1.5875, -1.0321, 0\right) \text{\AA}, 
\end{split}
\quad\hfill\quad
\begin{split}
\vec{r}_{13} \equiv \vec{r}_3 - \vec{r}_1 &= \left(1.4874, -2.9231, 0\right) \text{\AA}, \\
\vec{r}_{32} \equiv \vec{r}_2 - \vec{r}_3 &= \left(1.7877, 2.7497, 0\right) \text{\AA}, \\
\vec{r}_{21} \equiv \vec{r}_1 - \vec{r}_2 &= \left(-3.2752, 0.1734, 0\right) \text{\AA},
\end{split}
\end{equation}
where $\mathbf{R}_z(\theta)$ describes the rotation around the new $z$ axis by the angle $\theta$.  Here, we replaced Cu(1), Cu(2), and Cu(3) superscripts with 1, 2, and 3 to emphasize that we work in the new coordinate system associated with a single triangle. By applying $\mathbf{R}_{\Delta}$ from Eq.~(\ref{eq:r_delta}) to the DM vector in Eq.~(\ref{eq:d13_cryst}), we obtain the DM vectors in the $(x,y,z)$ frame:
\bea\label{eq:Dvectors1}
\mathbf{R}_{\Delta}\cdot\vec{D}_{\rm{T}}^{\text{Cu(1)-\text{Cu(3)}}} = \vec{D}_{\rm{T}}^{13} = -\vec{D}_{\rm{T}}^{31} &=& \left(-16.2, -8.6,  33.0\right) \text{K}\,.
\eea
The symmetrically equivalent DM vectors on other bonds of the triangle are:
\bea\label{eq:Dvectors2}
\vec{D}_{\rm{T}}^{21} = -\vec{D}_{\rm{T}}^{12} =& \mathbf{R}_z(\frac23\pi)\cdot\vec{D}_{\rm{T}}^{13} &= \left( 15.5, -9.7,  33.0\right) \text{K} \,,\\
\vec{D}_{\rm{T}}^{32} = -\vec{D}_{\rm{T}}^{23} =& \mathbf{R}_z(-\frac23\pi)\cdot\vec{D}_{\rm{T}}^{13} &= \left( 0.7,  18.3,  33.0\right) \text{K}\,.\label{eq:Dvectors3}
\eea
The orientation of $\vec{D}_{\rm{T}}$ vectors with respect to the constituents of the crystal structure is shown in Fig.~\ref{fig:dm_vecs}. The vectors are
perpendicular to the respective bonds and form $60^{\circ}$ angle with the line
connecting the midpoint of the respective bond and the third Cu atom. For
instance, for $\vec{D}_{\rm{T}}^{13}$, this line connects the midpoint of
$\vec{r}_{13}$ and Cu(2). The angle between each two of the three DM
vectors is close to $50^{\circ}$.

Finally, it is convenient to express $\vec{D}_{\rm{T}}$ in units of $J_{\rm{T}}\equiv\frac13\operatorname{Tr}\left[\mathbf{M}_{\rm{T}}\right]$ and decompose them into components perpendicular and parallel to the $z$ axis of a triangle, which coincides with the $c$ axis of the hexagonal unit cell in Fig.~\ref{fig:str}. We obtain:
\be\label{eq:dparallel_dperp}
D_{\rm{T}}^{\parallel}/J_{\rm{T}} = 0.493\,,~~
{D_{\rm{T}}^{\perp}}/{J_{\rm{T}}} = 0.274\,,~~
{D_{\rm{T}}^{\perp}}/{D_{\rm{T}}^{\parallel}} \simeq 1/\sqrt{3}\,. 
\ee

\section{The problem of a single antiferromagnetic spin-1/2 triangle with DM interactions}\label{sec:triangle}
We consider the problem of a single spin-1/2 triangle in the presence of AF exchange $J$, DM interactions and an external field ${\bf H}$. As in DiMACuS, the triangle features a threefold rotational axis along $z$ (perpendicular to the plane of the triangle), going through the center of the triangle. This transformation maps the individual spin operators as follows
\be
\renewcommand{\arraystretch}{1.3}
\begin{array}{l}
(S_1^x, S_1^y, S_1^z) \mapsto (-\frac{1}{2}S_2^x+\frac{\sqrt{3}}{2}S_2^y, -\frac{\sqrt{3}}{2}S_2^x-\frac{1}{2}S_2^y, S_2^z)\\
(S_2^x, S_2^y, S_2^z) \mapsto (-\frac{1}{2}S_3^x+\frac{\sqrt{3}}{2}S_3^y, -\frac{\sqrt{3}}{2}S_3^x-\frac{1}{2}S_3^y, S_3^z)\\
(S_3^x, S_3^y, S_3^z) \mapsto (-\frac{1}{2}S_1^x+\frac{\sqrt{3}}{2}S_1^y, -\frac{\sqrt{3}}{2}S_1^x-\frac{1}{2}S_1^y, S_1^z)
\end{array}
\ee

\subsection{Hamiltonian}
This problem is described by the Hamiltonian
\be\label{eq:TriangleHam}
\mc{H} = \mc{H}_J + \mc{H}_{\text{DM}}+\mc{H}_{\text{Z}}, 
\ee
where $\mc{H}_J$ comprises the Heisenberg interactions, $\mc{H}_{\text{DM}}$ the DM interactions, and $\mc{H}_{\text{Z}}$ the Zeeman coupling to the external field. 
The latter takes the usual form
\be
\mc{H}_{\text{Z}} = g \mu_B {\bf S}\cdot {\bf H}\,,
\ee
where we have assumed an isotropic $g$ factor. 
The Heisenberg terms take the form
\be\label{eq:HJ}
\mc{H}_J= J \left(\vec{S}_1\cdot\vec{S}_2+\vec{S}_2\cdot\vec{S}_3+\vec{S}_3\cdot\vec{S}_1\right)
=\frac{J}{2}\vec{S}^2-\frac{9}{8}J\,,
\ee
where ${\bf S}={\bf S}_1+{\bf S}_2+{\bf S}_3$ is the total spin. The DM terms take the form
\be
\mc{H}_{\text{DM}}= {\bf D}_{12}\cdot ({\bf S}_1\times{\bf S}_2)+{\bf D}_{23}\cdot ({\bf S}_2\times{\bf S}_3)+{\bf D}_{31}\cdot ({\bf S}_3\times{\bf S}_1)\,,
\ee
where the DM vectors ${\bf D}_{12}$, ${\bf D}_{23}$ and ${\bf D}_{31}$ are related to each other via the threefold rotational symmetry, namely 
\be
{\bf D}_{23} = {\bf R}\cdot {\bf D}_{12} ,~~{\bf D}_{31}={\bf R}\cdot {\bf D}_{23}  \,,
\ee
where 
\be
{\bf R}=\left(
\renewcommand{\arraystretch}{1.3}
\begin{array}{ccc}
-\frac{1}{2} & -\frac{\sqrt{3}}{2} & 0\\
\frac{\sqrt{3}}{2} & -\frac{1}{2} & 0\\
0 & 0 & 1
\end{array}
\right)\,.
\ee
More explicitly, denoting the three components of ${\bf D}_{12}$ by $a$, $b$ and $c$, we have
\be
{\bf D}_{12}=\left(a,b,c\right),~~~
{\bf D}_{23}=\left(-\frac{1}{2}a-\frac{\sqrt{3}}{2}b, \frac{\sqrt{3}}{2}a-\frac{1}{2}b, c\right),~~~
{\bf D}_{31}=\left(-\frac{1}{2}a+\frac{\sqrt{3}}{2}b, -\frac{\sqrt{3}}{2}a-\frac{1}{2}b, c\right)\,.
\ee
For later purposes, let us denote the terms arising from the out-of-plane components of the DM vectors by $\mc{H}_{\text{DM},z}$ and the terms arising from the in-plane components by $\mc{H}_{\text{DM},xy}$, namely
\be
\mc{H}_{\text{DM}}=\mc{H}_{\text{DM},z}+\mc{H}_{\text{DM},xy}\,.
\ee
The first term, which takes the form
\bea
\mc{H}_{\text{DM},z} &=& c \left(
S_1^x S_2^y - S_1^y S_2^x
+S_2^x S_3^y - S_2^y S_3^x
+S_3^x S_1^y - S_3^y S_1^x \right) \nonumber\\
&=&
-\frac{c}{2i} \left(
S_1^+ S_2^- - S_1^- S_2^+
+S_2^+ S_3^- - S_2^- S_3^+
+S_3^+ S_1^- - S_3^- S_1^+
\right),
\eea
plays a dominant role in DiMACuS, as we discuss in the main text.

\subsection{Convenient set of basis states and matrix representation of $\mc{H}$}
A convenient set $\mc{B}_1$ of basis states which diagonalizes $\mc{H}_J$. This set is labeled as $\{|S_{23},S,M\rangle\}$, where $S_{23}$ is the spin quantum number associated with ${\bf S}_{23}={\bf S}_2+{\bf S}_3$, $S$ is the total spin quantum number  (associated with ${\bf S}={\bf S}_1+{\bf S}_{23}$), and $M$ is its projection along the $z$ axis. The addition of three spins-1/2 gives two doublets and one quartet:
\be
\frac{1}{2}\otimes\frac{1}{2}\otimes\frac{1}{2}=\frac{1}{2}\otimes(0\oplus1)=\frac{1}{2}\oplus\frac{1}{2}\oplus\frac{3}{2}\,.
\ee
More explicitly, the quantum number $S_{23}$ can be either 0 or 1. For $S_{23}=0$, we get the $S=1/2$ doublet
\bea
|0,1/2,1/2\rangle = \frac{1}{2}\left( |\!\upa \upa\dna\rangle - |\!\upa\dna\upa\rangle \right),~~~|0,1/2,-1/2\rangle = \frac{1}{2}\left( |\!\dna \upa\dna\rangle - |\!\dna\dna\upa\rangle\right)\,.
\eea
For $S_{23}=1$, the total spin can be either $1/2$ or $3/2$. 
The former consists of the states
\be
|1,1/2,1/2\rangle=\frac{1}{\sqrt{6}} \left( 2 |\dna\upa\upa\rangle-|\upa\dna\upa\rangle- |\upa\upa\dna\rangle\right)\,,~~~
|1,1/2,-1/2\rangle=\frac{1}{\sqrt{6}} \left( 2 |\upa\dna\dna\rangle-|\dna\upa\dna\rangle- |\dna\dna\upa\rangle\right)\,,
\ee
whereas the $S=3/2$ quartet consists of
\be\label{eq:quartet}
\renewcommand{\arraystretch}{1.3}
\begin{array}{l}
|1,3/2,3/2\rangle=|\!\upa\upa\upa\rangle\\
|1,3/2,1/2\rangle = \frac{1}{\sqrt{3}} \left( |\!\dna\upa\upa\rangle + |\!\upa\dna\upa\rangle+|\!\upa\upa\dna\rangle \right) \\
|1,3/2,-1/2\rangle = \frac{1}{\sqrt{3}} \left( |\!\upa\dna\dna\rangle + |\!\dna\upa\dna\rangle+|\!\dna\dna\upa\rangle \right)\,,\\
|1,3/2,-3/2\rangle=|\!\dna\dna\dna\rangle\,.
\end{array}
\ee
According to Eq.~(\ref{eq:HJ}), the $\{|S_{23},S,M\rangle\}$ basis is an eigenbasis of $\mc{H}_J$, with
\be
\mc{H}_J |S_{23},S,M\rangle = \left(\frac{J}{2} S(S+1) -\frac{9}{8}J\right)|S_{23},S,M\rangle\,.
\ee
The two doublets are degenerate with energy $-3J/4$ 
and the quartet has energy $3J/4$. 

Now, let us examine the symmetry of these states under the threefold rotation which maps the spin indices (123) to (312). In general, the eigenvalues of $C_3$ can be labeled as $e^{i \frac{2\pi}{3}\ell}$, where $\ell=0$, $1$ and $-1$. The quartet belongs to the sector $\ell=0$, since $C_3 |1,3/2,M\rangle=|1,3/2,M\rangle$ for any $M$. The two doublets are not eigenstates of $C_3$ but we can form  linear combinations which are. These combinations can be labeled as $|\ell,\sigma\rangle$ with $\ell=\pm 1$ and $\sigma=\pm1/2$ (or $\upa$ and $\dna$), and are given by: 
\be\label{eq:ellsigma}
|\ell,\sigma\rangle = \frac{1}{\sqrt{3}} \left( |-\sigma,\sigma,\sigma\rangle+e^{-i\frac{2\pi}{3}\ell} |\sigma,-\sigma,\sigma\rangle+e^{i\frac{2\pi}{3}\ell} |\sigma,\sigma,-\sigma\rangle\right)\,,
\ee
with 
\bea
C_3~|\ell,\sigma\rangle = e^{i\frac{2\pi}{3} \ell} ~|\ell,\sigma\rangle \,.
\eea
Note that the $M=\pm1/2$ members of the quartet [second and third line of (\ref{eq:quartet})] are also of the form of Eq.~(\ref{eq:ellsigma}), but with $\ell=0$. Altogether, we have the following new set $\mc{B}_2$ of basis states:
\be
\mc{B}_2  =\left\{
|\ell\!=\!1,\upa\rangle,~
|\ell\!=\!-1,\dna\rangle,~
|\ell\!=\!-1,\upa\rangle,~
|\ell\!=\!1,\dna\rangle,~
|\!\upa\upa\upa\rangle,~
|\ell\!=\!0,\upa\rangle,~ 
|\ell\!=\!0,\dna\rangle,~ 
|\!\dna\dna\dna\rangle
\right\}\,.
\ee
In this basis, the Hamiltonian in the presence of a field 
\small
\be
{\bf H}={\bf h}/(g\mu_B),~~{\bf h}=h_\perp (\cos\phi\hat{\bf x}+\sin\phi\hat{\bf y})+h_z\hat{\bf z}\,,
\ee
\normalsize
takes the form
\small
\be\label{eq:HmatrixB2}
\mc{H}\!=\!\frac{1}{2}\!\left(
\renewcommand{\arraystretch}{1.2}
\begin{array}{cc|cc|cccc}
-\frac{3J}{2}\!-\!\sqrt{3}c\!+\!h_z&0&0&{\cred-h_\perp e^{-i\phi}}&0&0&{\cbl\gamma^\ast}&0\\
0&-\frac{3J}{2}\!-\!\sqrt{3}c\!-\!h_z&{\cred-h_\perp e^{i\phi}}&0&0&{\cbl-\gamma}&0&0\\
\hline
0&{\cred-h_\perp e^{-i\phi}}&-\frac{3J}{2}\!+\!\sqrt{3}c\!+\!h_z&0&{\cbl\sqrt{3}\gamma}&0&0&0\\
{\cred-h_\perp e^{i\phi}}&0&0&-\frac{3J}{2}\!+\!\sqrt{3}c\!-\!h_z&0&0&0&{\cbl-\sqrt{3}\gamma^\ast}\\
\hline
0&0&{\cbl\sqrt{3}\gamma^\ast}&0&\frac{3J}{2}\!+\!3h_z&{\cred\sqrt{3}h_\perp e^{-i\phi}}&0&0\\
0&{\cbl-\gamma^\ast}&0&0&{\cred\sqrt{3}h_\perp e^{i\phi}}&\frac{3J}{2}\!+\!h_z&{\cred 2 h_\perp e^{-i\phi}}&0\\
{\cbl\gamma}&0&0&0&0&{\cred 2 h_\perp e^{i\phi}}&\frac{3J}{2}\!-\!h_z&{\cred\sqrt{3}h_\perp e^{-i\phi}}\\
0&0&0&{\cbl-\sqrt{3}\gamma}&0&0&{\cred\sqrt{3}h_\perp e^{i\phi}}&\frac{3J}{2}\!-\!3h_z\\
\end{array}\right)_{\!\!\!\!\mc{B}_2}.~~~~~~~
\ee
\normalsize
where $\gamma \equiv \frac{\sqrt{3}}{2}e^{i2\pi/3}(a+i b)$. 

\subsection{Energy spectrum}
The field dependence of the energy spectrum is shown in Fig.~\ref{fig:Evsh} for ${\bf h}\!\parallel\!{\bf z}$ (panel a) and ${\bf h}\!\parallel\!{\bf x}$-axis (panel b), for a representative set of DM parameters $(c,d_\perp\equiv\sqrt{a^2+b^2})\!=\!(0.2J,0)$ (solid lines)  and $(0.2J,0.2J)$ (dotted lines).

\begin{figure}[!h]
\includegraphics[width=0.49\linewidth]{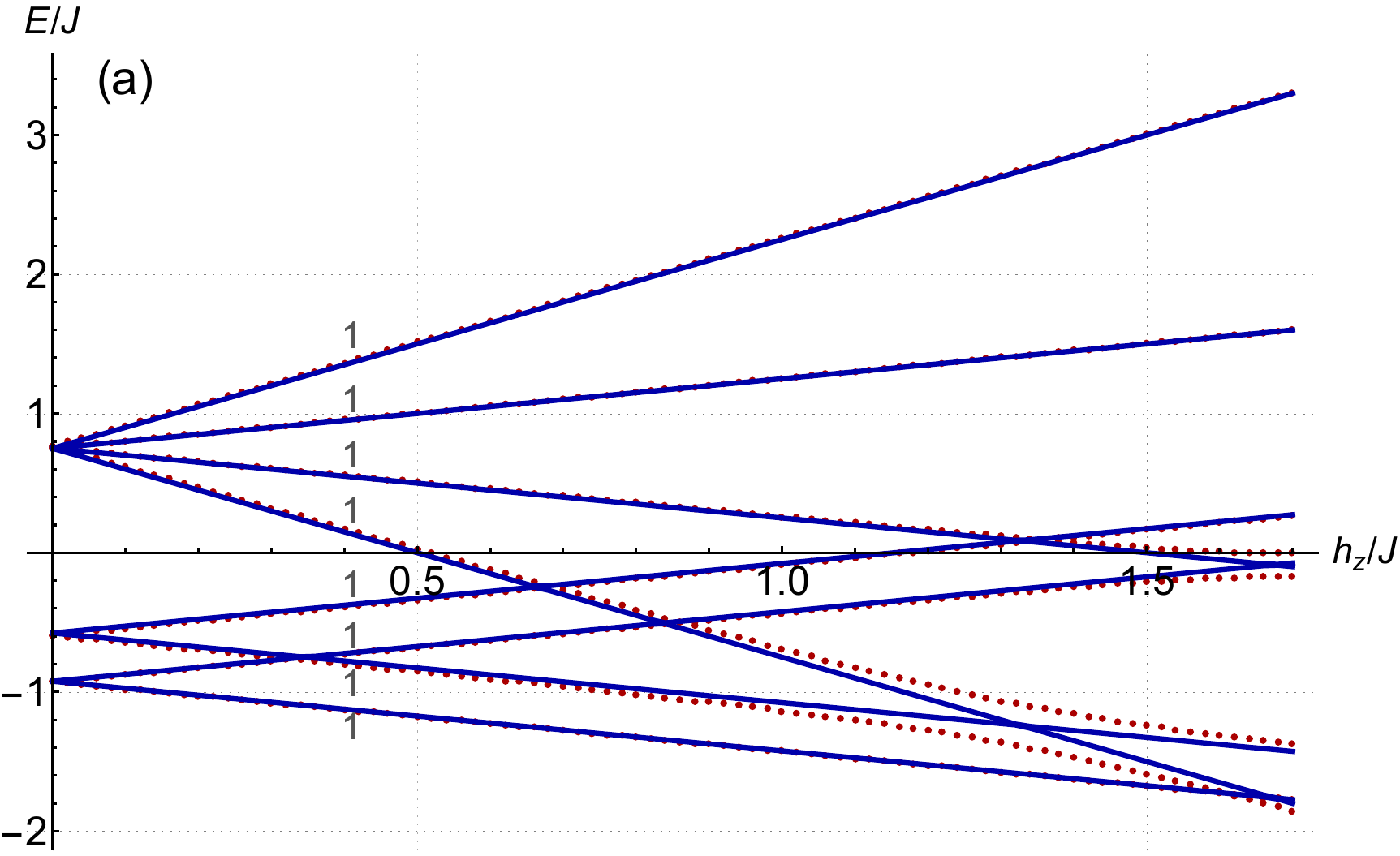}
\includegraphics[width=0.49\linewidth]{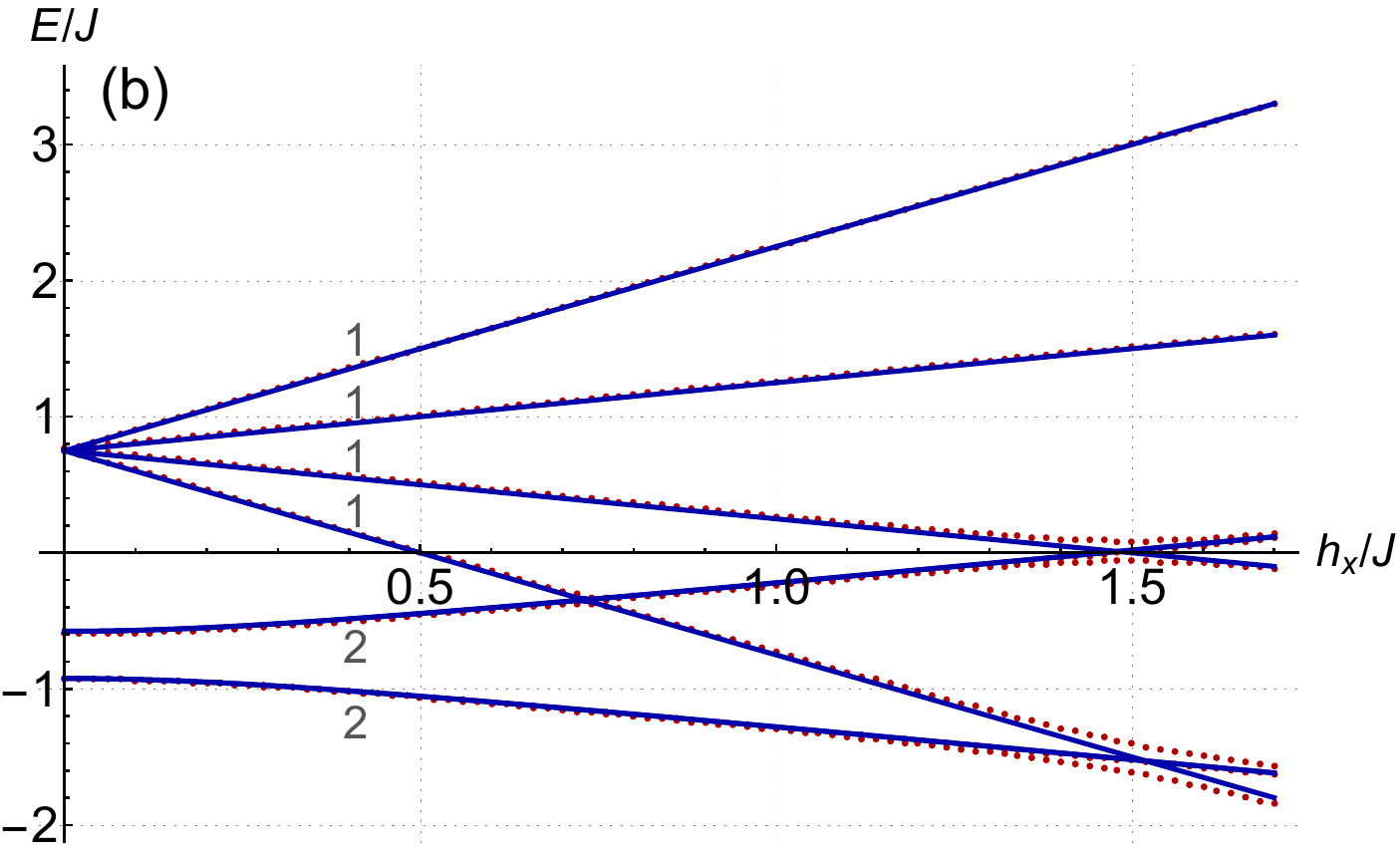}
\caption{Evolution of the energy spectrum of $\mc{H}_{J}\!+\!\mc{H}_{\text{DM}}\!+\!{\bf h}\cdot {\bf S}$, where ${\bf h}\!=\!g\mu_B{\bf H}$, for ${\bf h}\!\parallel\!{\bf z}$ (a) and ${\bf h}\!\parallel\!{\bf x}$-axis (b). The DM parameters $c$ and $d_\perp\equiv\sqrt{a^2+b^2}$ are $(c,d_\perp)=(0.2J,0)$ for the blue solid lines, and $(0.2J,0.2J)$ for the red dotted lines. The number labels give the degeneracy of each energy level.}\label{fig:Evsh}
\end{figure}

\noindent
The main features of Fig.~\ref{fig:Evsh} can be understood by examining the general structure of the matrix of Eq.~(\ref{eq:HmatrixB2}): 
\begin{itemize}[leftmargin=0.25in]
\setlength\itemsep{0.02in}

\item $\mc{B}_2$ is an eigenbasis of $\mc{H}_{\text{DM},z}$. This is because $\mc{H}_{\text{DM},z}$ is invariant under: i) $C_3$ and thus does not couple states with different $\ell$; ii) $U(1)$ rotations around the $z$ axis in spin space and thus does not couple states with different $S_z$. 

\item At zero field, $\mc{H}_{\text{DM},z}$ lifts the degeneracy of the two doublet states of $\mc{H}_J$ and gives rise to one Kramers doublet $\{|\ell\!=\!1,\upa\rangle,|\ell\!=\!-1,\dna\rangle\}$ with $E\!=\!-3J/4\!-\!\sqrt{3}c/2$ and a second Kramers doublet $\{|\ell\!=\!-1,\upa\rangle,|\ell\!=\!1,\dna\rangle\}$ with $E\!=\!-3J/4\!+\!\sqrt{3}c/2$. Essentially, $c$ locks the quantum number $\ell$ to the value of $\sigma$.

\item Unlike $\mc{H}_{\text{DM},z}$, $\mc{H}_{\text{DM},xy}$ [matrix elements highlighted in blue color in Eq.~(\ref{eq:HmatrixB2})] is not diagonal in the $\mc{B}_2$ basis. More importantly, $\mc{H}_{\text{DM},xy}$ does not couple the two doublets, but only connects the doublets with the high-energy quartet. As a result, the leading correction to the energy eigenvalues from the transverse DM components $a$ and $b$ is second order in $d_\perp/J$, where $d_\perp\!\equiv\!\sqrt{a^2+b^2}$. In other words, the transverse DM components have little influence on the spectrum (see difference between solid and dotted lines in Fig.~\ref{fig:Evsh}), and are therefore not as significant as the $z$ components. In particular, $\mc{H}_{\text{DM},xy}$ does not influence the quartet, as long as we are away from level anticrossings.

\item $\mc{H}_{\text{DM},z}$ does not affect the quartet. So, in the absence of $\mc{H}_{\text{DM},xy}$, the quartet will show the usual Zeeman splitting for any field direction, see solid lines in Fig.~\ref{fig:Evsh}. A nonzero $d_\perp$ gives rise to an excited level antocrossing at high fields, see dotted lines in Fig.~\ref{fig:Evsh}.

\item For a magnetic field along the $z$-axis, the Zeeman coupling is diagonal in the basis $\mc{B}_2$ because it conserves $S_z$ and $\ell$ (the Zeeman term is invariant under $C_3$). Hence, a field along $z$ gives rise to the usual Zeeman splitting of each separate multiplet, see Fig.~\ref{fig:Evsh}\,(a).

\item By contrast, for in-plane fields, the Zeeman coupling [matrix elements highlighted in red color in Eq.~(\ref{eq:HmatrixB2})] connects states with the same $\ell$ and $\Delta S_z=\pm1$. As a result, the in-plane magnetic field gives rise to: i) a level-repulsion of the two doublets of $\mc{H}_{J}+\mc{H}_{\text{DM},z}$, without lifting the two-fold degeneracy of each doublet, and, ii) the usual Zeeman splitting of the quartet. Figure~\ref{fig:Evsh}\,(b) illustrates these aspects.

\end{itemize}

\noindent We can also deduce the following analytical results for the energy spectrum in special cases:

\begin{itemize}[leftmargin=0.25in]
\setlength\itemsep{0.02in}

\item For fields along the $z$-axis and $d_\perp\!=\!0$, the energies are given by the diagonal elements of the matrix of Eq.~(\ref{eq:HmatrixB2}), namely
\small
\be\label{eq:EdcHz}
\text{spectrum of}~\mc{H}_{J}\!+\!\mc{H}_{\text{DM},z}\!+\!h_z S^z:~
\boxed{
-\frac{3J}{4}-\frac{\sqrt{3}|c|}{2}\pm\frac{h_z}{2},~~
-\frac{3J}{4}+\frac{\sqrt{3}|c|}{2}\pm\frac{h_z}{2},~~
\frac{3J}{4}\pm\frac{3h_z}{2},~~
\frac{3J}{4}\pm\frac{h_z}{2}
}\,, 
\ee
\normalsize

\item For fields along the $z$-axis and $d_\perp\!\neq\!0$, the energies are given by
\small
\be\label{eq:EHz}
\!\!\!\!\!\text{spectrum of}~\mc{H}_{J}+\mc{H}_{\text{DM}}+h_z S^z:
\boxed{-\frac{\sqrt{3}}{4} c \!+\! \frac{\zeta_1}{4}\sqrt{3 d_\perp^2 \!+\! (\sqrt{3} c \!+\!3J \!+\!\zeta_2 2 h_z)^2 }, 
\frac{\sqrt{3}}{4} c \!+\!h_z \!+\! \frac{\zeta_1}{4}\sqrt{9 d_\perp^2 \!+\! (\sqrt{3} c\!-\!3J\!+\!\zeta_2 2 h_z)^2 }
},
\ee
\normalsize
where $\zeta_1=\pm1$ and $\zeta_2=\pm1$.

\item For fields along the $x$-axis (the results are the same for any direction in the $xy$-plane) and $d_\perp\!=\!0$,  the energies are given by 
\small
\be\label{eq:EdcHx}
\text{spectrum of}~\mc{H}_{J}\!+\!\mc{H}_{\text{DM},z}\!+\!h_xS^x:~
\boxed{
-\frac{3J}{4}\pm \frac{1}{2}\sqrt{3c^2+h_x^2}~(\text{each two-fold}),~~
\frac{3J}{4}\pm \frac{3h_x}{2},~~
\frac{3J}{4}\pm \frac{h_x}{2}
}\,.
\ee
\normalsize

\item For fields along the $x$-axis and $d_\perp\!\neq\!0$, we can obtain the energies up to second order in $d_\perp$ using perturbation theory (valid away from level-anticrossings). We find
\small
\be\label{eq:EHx}
\text{spectrum of}~\mc{H}_{J}\!+\!\mc{H}_{\text{DM}}\!+\!h_xS^x:~
\boxed{
-\frac{3J}{4}\pm \frac{1}{2}\sqrt{3c^2+h_x^2}+\delta E^{(2)}_{\pm} ~ (\text{each two-fold}),~~
\frac{3J}{4}\pm \frac{3h_x}{2},~~
\frac{3J}{4}\pm \frac{h_x}{2}
}\,.
\ee
\normalsize
where
\small
\be
\delta E^{(2)}_{+}=-\frac{|\gamma|^2}{32}\frac{(3 \sqrt{3} c + h_x + 3 \sqrt{3 c^2 + h_x^2})^2}{(3 c^2 + h_x^2 + \sqrt{3} c \sqrt{3 c^2 + h_x^2}) (3 J+h_x-\sqrt{3 c^2 + h_x^2})},~~
\delta E^{(2)}_{-}=\frac{|\gamma|^2}{16}
\frac{4 \sqrt{3} c + 3 h_x - 5 \sqrt{3 c^2 + h_x^2}}{\sqrt{3 c^2 + h_x^2}~(3J+h_x+\sqrt{3 c^2 + h_x^2})}\,.
\ee
\normalsize
It follows that, to leading order in $d_\perp$: 
i) the zero-field energy of the quartet is unaffected by $d_\perp$, and, ii) the zero-field splitting between the two lowest doublets becomes
\be
\boxed{
\Delta =  \sqrt{3}|c| + \frac{J}{32} \left(\frac{d_\perp}{J}\right)^2 \frac{12+5 \sqrt{3}\frac{c}{J}}{3-\left(\frac{c}{J}\right)^2}
}\,.
\ee
\end{itemize}

\subsection{Magnetization process for $d_\perp=0$}
In the following we shall disregard the in-plane components of the DM vectors. The magnetization can be obtained using the expression
\be\label{eq:mvec}
{\bf m}=-\sum_n P_n \frac{\partial E_n}{\partial {\bf H}}\,,
\ee
where $\{E_n, n=1-8\}$ are the energy eigenvalues, $P_n=e^{-\beta E_n}/Z$ are the corresponding Boltzmann probabilities, $\beta\!=\!\frac{1}{k_BT}$ and $Z\!=\!\sum_n e^{-\beta E_n}$ is the partition function. We find:

\begin{itemize}[leftmargin=0.25in]
\setlength\itemsep{0.05in}

\item {\bf Magnetization process for ${\bf h}\!\parallel\!{\bf z}$:} Using Eqs.~(\ref{eq:EdcHz}) and (\ref{eq:mvec}) leads to the analytical expression
\small
\be
\boxed{
\frac{m_z}{g\mu_B}=\frac{1}{2}\tanh\left(\frac{\beta h_z}{2}\right)\left[
1+4\frac{\cosh^2(\beta h_z/2)}{\cosh(\beta h_z)+e^{3\beta J/2}\cosh(\sqrt{3}\beta |c|/2)}
\right]
}\,.
\ee
\normalsize
The evolution of $m_z$ with $h_z/J$ for $k_BT/J=0.005$ and $|c|/J=0$ (black), 0.1 (blue) and 0.2 (red) is shown in Fig.~\ref{fig:DMc-mvsh}\,(a). First of all, the results show a clear 1/2 magnetization plateau. Second, the magnetization shows very little dependence on $c$ (invisible in the graph) up to the saturation field $h^\ast_{z}$, whose value seems to be the only aspect that changes with $c$. Specifically, $h^\ast_{z}$ appears to increase linearly with $c$. Analytically, we find (at $T=0$)
\be
\boxed{h^\ast_{z}=3J/2+\sqrt{3}|c|/2}\,.
\ee

\item {\bf Magnetization process for ${\bf h}\!\perp\!{\bf z}$:} Similarly, we can obtain the magnetization $m_x$ for for ${\bf h}\!\perp\!{\bf z}$ and $d_\perp\!=\!0$ using Eqs.~(\ref{eq:EdcHx}) and (\ref{eq:mvec}), but the resulting analytical expression for $m_x$ is too lengthy to write down explicitly. However, we can obtain a simpler formula for the zero-temperature magnetization, where Eq.~(\ref{eq:mvec}) reduces to
\be\label{eq:mxT0}
T=0:~~
\frac{m_x}{g\mu_B}
=-\frac{\partial}{\partial h_x} E_1 \,,
\ee
where $E_1$ is the ground state energy, 
\be
E_1=\Bigg\{
\begin{array}{ll}
-\frac{3J}{4}-\frac{1}{2}\sqrt{3c^2+h_x^2}, & \text{if}~|h_x|\leq h_x^\ast\\
+\frac{3J}{4}-\frac{3}{2}h_x, & \text{if}~ |h_x|\geq h_x^\ast
\end{array}
\ee
and $h_x^\ast$ is the saturation field. We find: 
\be
\!\!\boxed{
\frac{m_x}{g\mu_B}=\Bigg\{
\begin{array}{ll}
\frac{1}{2} \frac{h_x}{\sqrt{3c^2+h_x^2}}, & \text{if}~0 \leq h_x\leq h_x^\ast\\
\frac{3}{2}, & \text{if}~ h_x\geq h_x^\ast
\end{array}
}~\text{and}~
\boxed{
h_x^\ast=\frac{1}{8}\left(9J+\sqrt{9J^2+24c^2}\right)
=J\big[\frac{3}{2}+\frac{1}{2}(c/J)^2+\mc{O}(c/J)^4\big]
}\,.
\ee
Note that the saturation field $h_x^\ast$ scales quadratically with $c$ for small $c/J$, and therefore is not influenced from $c$ as strongly as $h_z^\ast$. This is also demonstrated in Fig.~\ref{fig:DMc-mvsh}.

strongly on $c$ quadratically quaThe size of the magnetization jump at $h_x^\ast$ is given by
\be
\frac{\delta m_x}{g\mu_B}=
\frac{3}{2} - \frac{9 + \sqrt{9 + 244 (c/J)^2}}{16\sqrt{3 (c/J)^2 + \frac{1}{64} (9 + \sqrt{9+ 244 (c/J)^2})^2}}=1+\frac{(c/J)^2}{3}+\mc{O}((c/J)^4)\,.
\ee
The low-$T$ magnetization process for a field applied along the $x$ axis is shown in Fig.~\ref{fig:DMc-mvsh}\,(b) for $|c|/J=0$, 0.1 and 0.2. The response is qualitatively different from Fig.~\ref{fig:DMc-mvsh}\,(a). Most importantly, the flat plateau disappears for nonzero $c$. Second, it is the low-field regime of $m_x$ which is affected by $c$, and not the high-field regime, which is opposite to what happens in Fig.~\ref{fig:DMc-mvsh}\,(a). Third, the slope of $m_x$ at low fields does not diverge (as for the case of $c\!=\!0$ or for fields along $z$) and decreases with increasing $c$, see next subsection.

\end{itemize}

\begin{figure}[!t]
\includegraphics[width=0.49\linewidth]{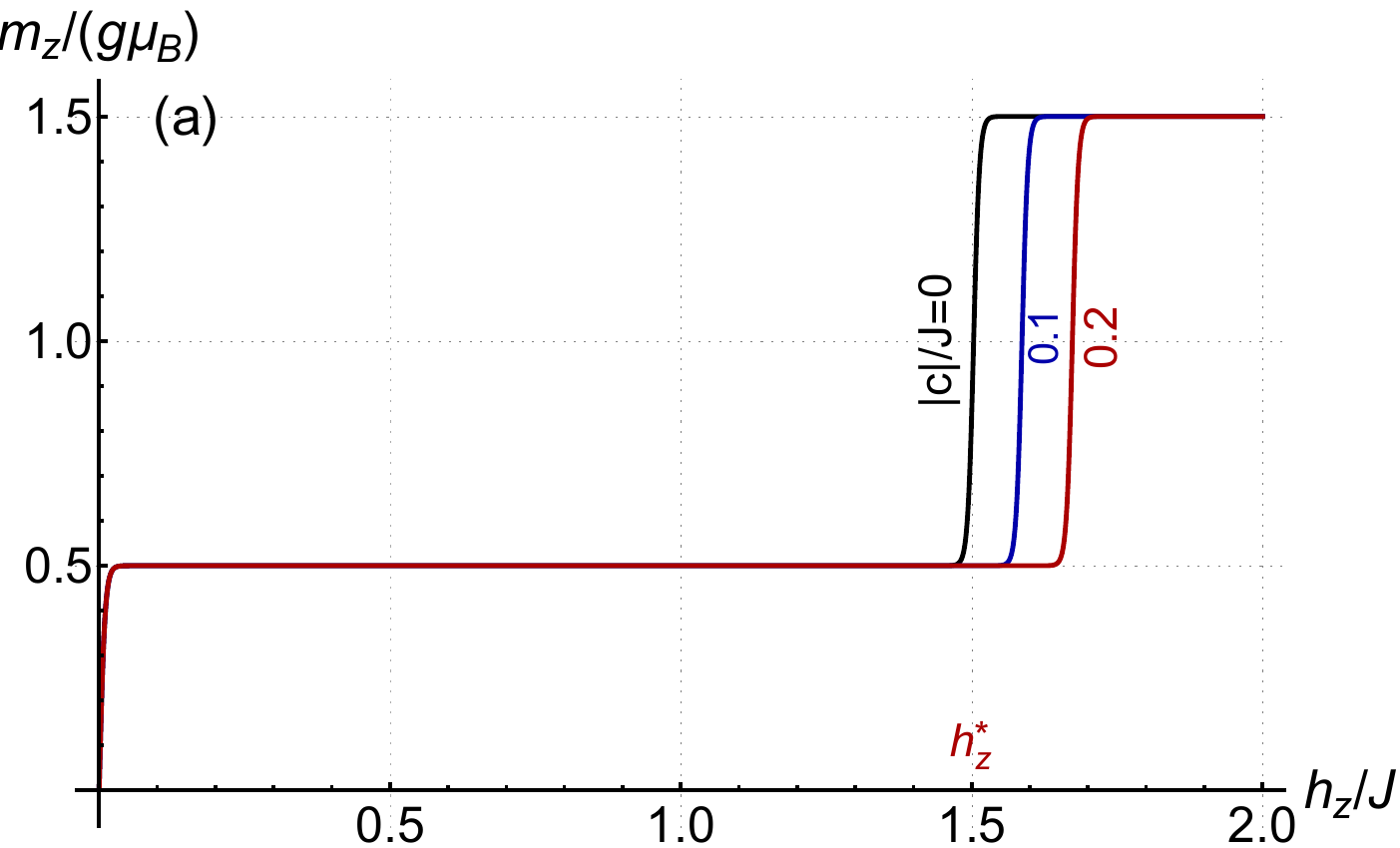}
\includegraphics[width=0.49\linewidth]{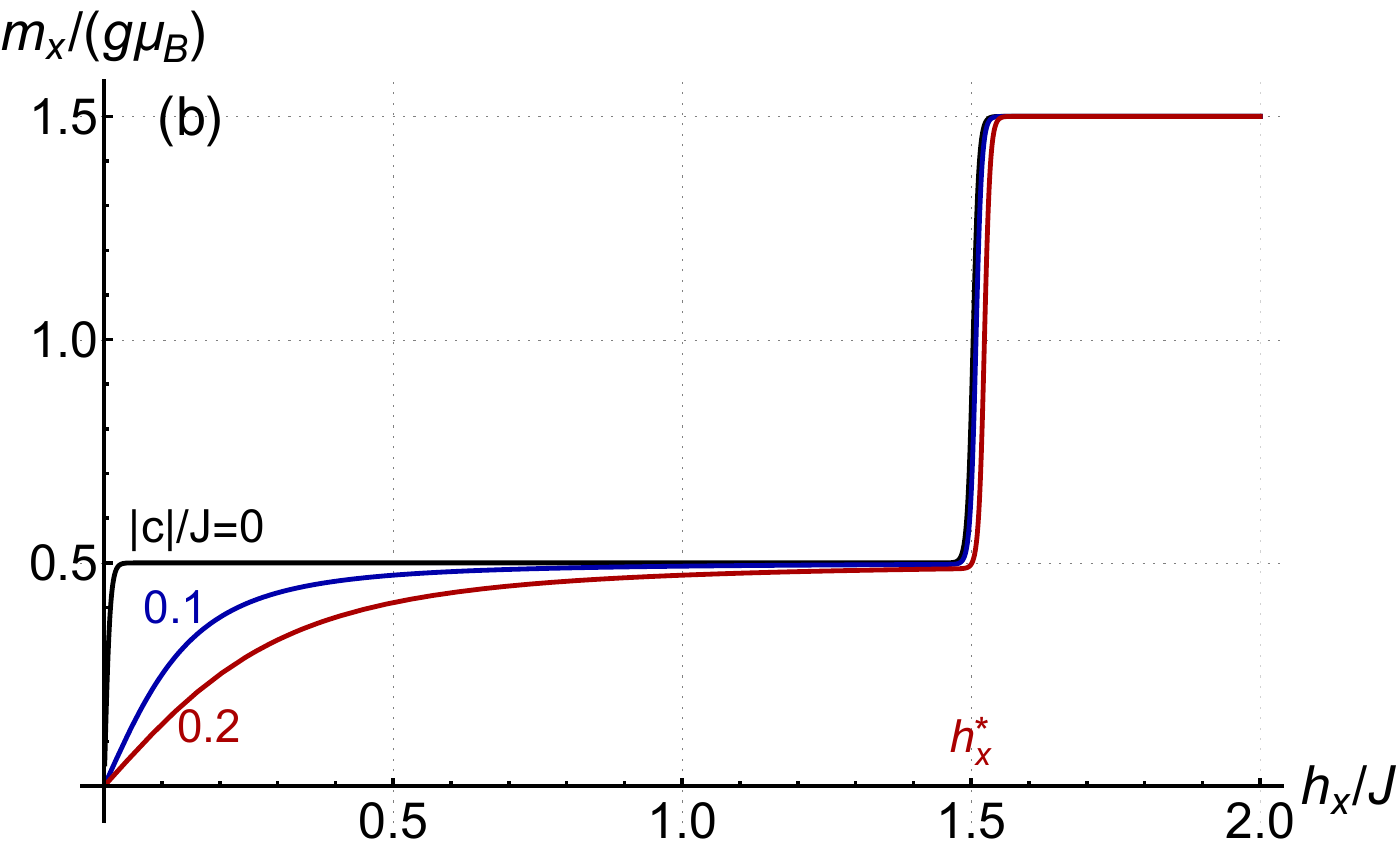}
\caption{Low-$T$ magnetization process of a spin-1/2 triangle described by $\mc{H}_{J}\!+\!\mc{H}_{\text{DM},z}\!+\!{\bf h}\cdot {\bf S}$, for ${\bf h}\!\parallel\!{\bf z}$ (a) and ${\bf h}\!\parallel\!{\bf x}$-axis (b). Here $k_B T=0.005J$, ${\bf h}=g\mu_B {\bf H}$ and the three curves correspond to $|c|/J=0$ (black), 0.1 (blue) and 0.2 (red).}\label{fig:DMc-mvsh}
\end{figure}

\subsection{Magnetic susceptibility for $d_\perp=0$}
We can find an analytical expression for the susceptibility using the Van Vleck formula
\be\label{eq:VV}
\chi_{\alpha\alpha'} =\lim_{H\to0} \sum_n P_n \left( \beta \frac{\partial E_n}{\partial H_\alpha} \frac{\partial E_n}{\partial H_{\alpha'}}-\frac{\partial^2 E_n}{\partial H_\alpha \partial H_{\alpha'}}\right)\,,
\ee
where $\alpha$ and $\alpha'$ are Cartesian components and $\chi_{\alpha\alpha'}=\lim_{H\to 0}\frac{M_\alpha}{H_{\alpha'}}$. Due to symmetry, the susceptibility tensor is diagonal and, furthermore, $\chi_{xx}=\chi_{yy}$. In particular, we find (disregarding again the in-plane components of the DM vectors):

\begin{itemize}[leftmargin=0.25in]
\setlength\itemsep{0.05in}

\item {\bf Longitudinal susceptibility:}  Using Eq.~(\ref{eq:VV}) and the expressions for the energy eigenvalues given in Eq.~(\ref{eq:EdcHz}), gives
\be
\label{eq:chizz}
\boxed{
\frac{\chi_{zz}}{(g\mu_B)^2}=\beta J\left(\frac{1}{4}+\frac{1}{1+e^{3\beta J/2} \cosh(\sqrt{3}\beta |c|/2)}\right)
}\,.
\ee
The evolution of $\chi_{zz}$ with $k_B T/J$ for $|c|/J=0.1$ and $0.2$ is shown in Fig.~\ref{fig:DMc-chi-vs-H}\,(a). The curves fall almost on top of each other (they start to deviate slightly at high $T$ for larger values of $|c|/J$). So, $\chi_{zz}$ is essentially unaffected by $c$, which is consistent with the agreement between the three curves of Fig.~\ref{fig:DMc-mvsh}\,(a) at low fields. 
Moreover, at low temperatures, 
\be\label{eq:lowT-chizz}
\boxed{\text{low}~T:~~ \frac{\chi_{zz}}{(g\mu_B)^2} \to \frac{\beta J}{4}}\,,
\ee
which is the same expression with that of a spin-1/2 object (Curie's law). 
At high temperatures, 
\be\label{eq:highT-chizz}
\boxed{
\text{high}~T:~~ \frac{\chi_{zz}}{(g\mu_B)^2} \to \frac{3}{4}(\beta J)-\frac{3}{8}(\beta J)^2 -\frac{3}{32}\beta^3 J c^2  + \mc{O}(\beta^4)
}\,,
\ee
So, the two leading terms of the high-$T$ expansion of $\chi_{zz}$ are unaffected by the DM interactions. These leading terms give a Curie-Weiss temperature $k_B\Theta=-J/2$, the same as in the case without DM interactions. This is a general result that is related to the fact that the DM interactions are antisymmetric (see, e.g., appendix of Ref.~[\onlinecite{sm:panther2023frustration}]).

\item {\bf Transverse susceptibility:} In a similar way, using Eq.~(\ref{eq:VV}) and the energy expressions given in Eq.~(\ref{eq:EdcHx}), gives
\be
\label{eq:chixx}
\boxed{
\frac{\chi_{xx}}{(g\mu_B)^2} = 
\frac{J}{2\sqrt{3}|c|} \frac{1-e^{-\sqrt{3}\beta|c|}}{1+2e^{-3\beta J/2}e^{-\sqrt{3}\beta|c|/2}+e^{-\sqrt{3}\beta|c|}}+\frac{5\beta J}{4}\frac{1}{1+e^{3\beta J/2}\cosh(\beta \sqrt{3}|c|/2)}
}\,.
\ee
The evolution of $\chi_{xx}$ with $k_B T/J$ for $|c|/J=0$, $0.1$ and $0.2$ is shown in Fig.~\ref{fig:DMc-chi-vs-H}\,(b). The results show a qualitatively different behaviour from the curves shown in Fig.~\ref{fig:DMc-chi-vs-H}\,(a), which stems from the corresponding difference in the low-field magnetization process shown in the two panels of Fig.~\ref{fig:DMc-mvsh}. 
Unlike $\chi_{zz}$, $\chi_{xx}$ does not diverge as $1/T$ at low $T$, but saturates to a $T$-independent value that is inversely proportional to the $z$ component of the DM vectors. Specifically, 
\be\label{eq:lowT-chixx}
\boxed{\text{low}~T:~~\frac{\chi_{xx}}{(g\mu_B)^2}\to \frac{J}{2\sqrt{3}|c|}}\,.
\ee
At high temperatures, 
\be\label{eq:highT-chixx}
\boxed{
\text{high}~T:~~ \frac{\chi_{xx}}{(g\mu_B)^2} \to \frac{3}{4}(\beta J)-\frac{3}{8}(\beta J)^2 -\frac{1}{8}\beta^3 J c^2  + \mc{O}(\beta^4)
}\,,
\ee
The two leading terms are again unaffected by the DM interactions, and, moreover, coincide with the two leading terms of the high-$T$ expansion of $\chi_{zz}$, see Eq.~(\ref{eq:highT-chizz}). 
The third term in the expansion differs by that of Eq.~(\ref{eq:highT-chizz}) by a factor of $4/3$, which reflects the general trend of the $\chi_{xx}$ curves (corresponding to $c\neq0$) to lie below the $\chi_{zz}$ curve, see Fig.~\ref{fig:DMc-chi-vs-H}.

\end{itemize}

\begin{figure}[!h]
\includegraphics[width=0.49\linewidth]{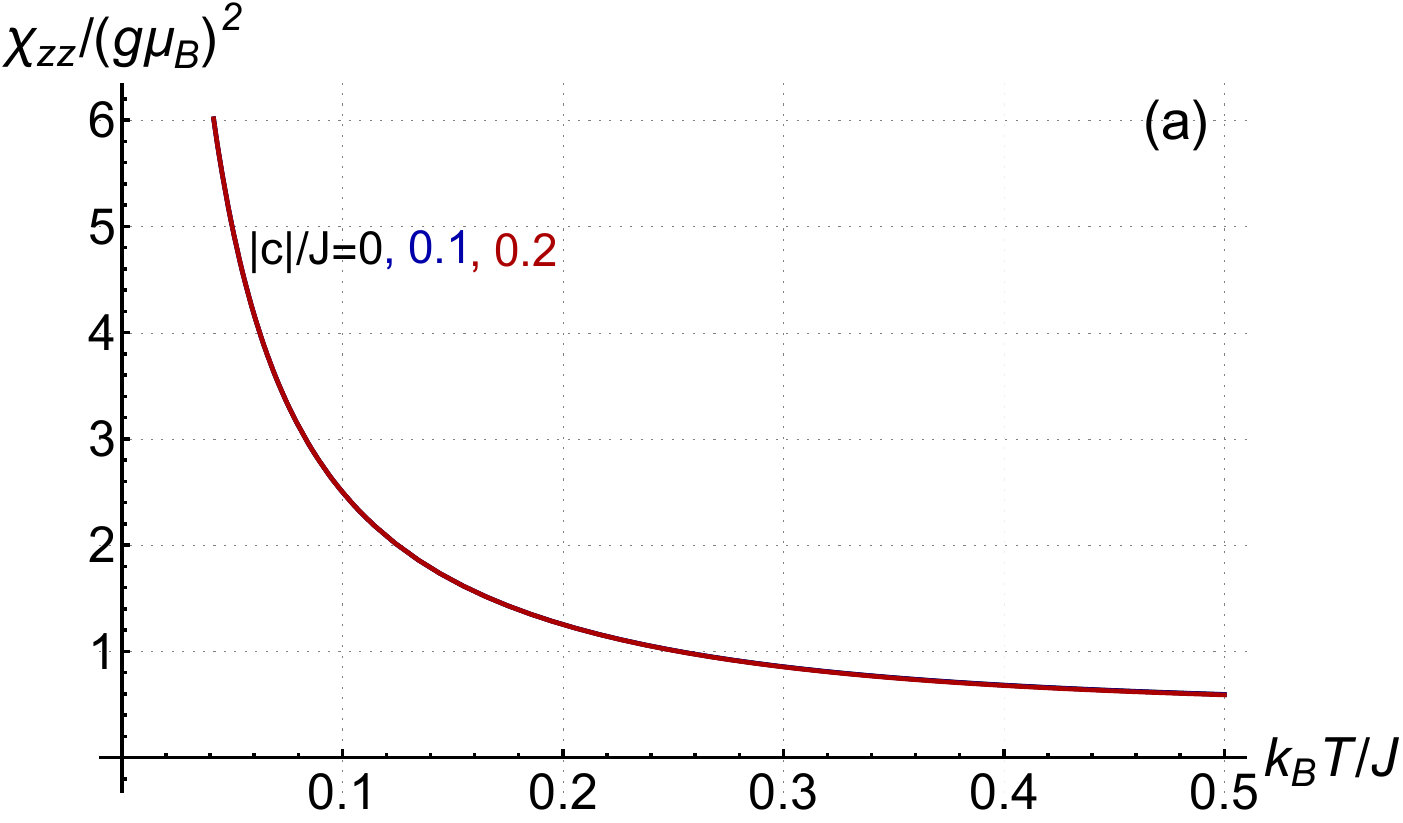}
\includegraphics[width=0.49\linewidth]{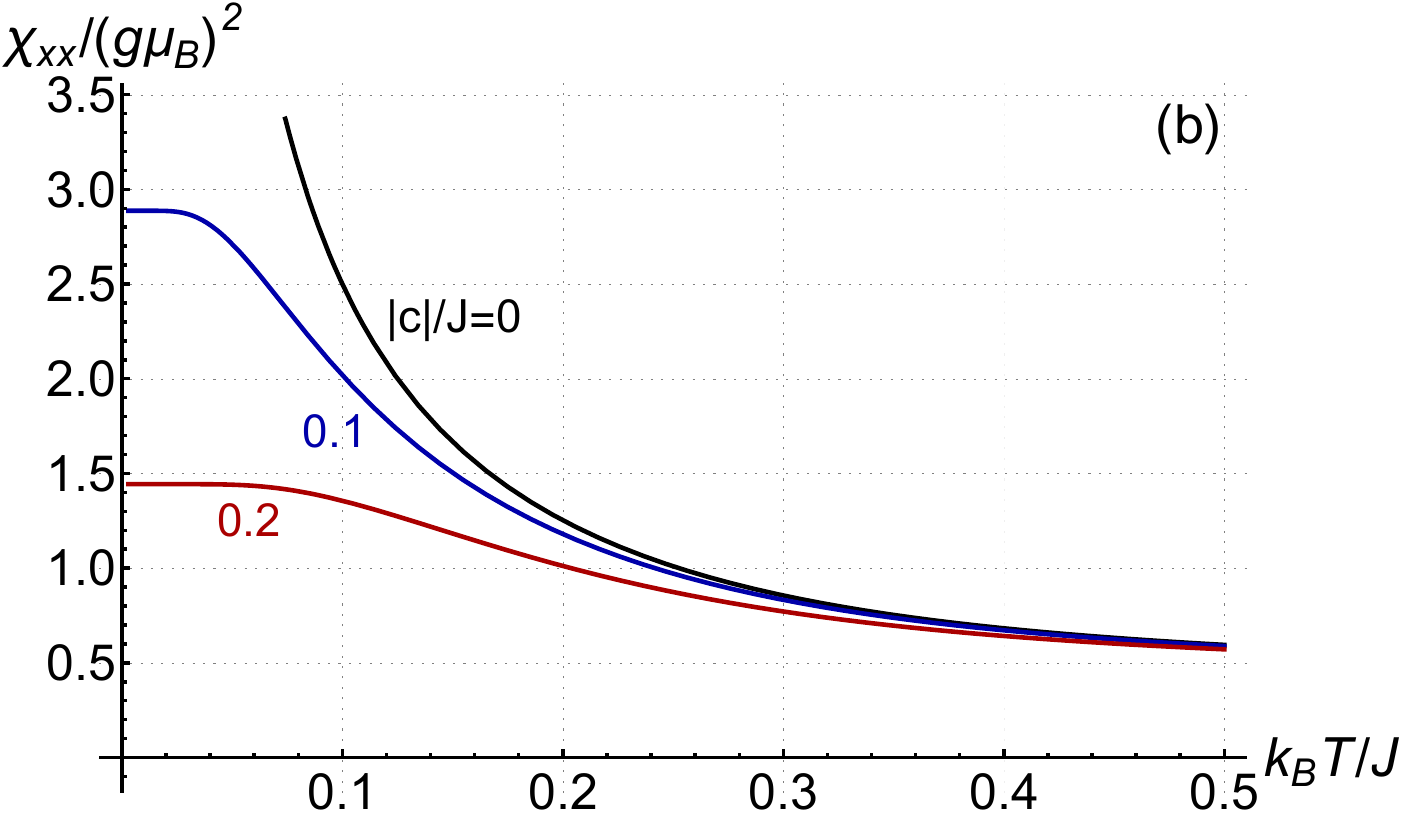}
\caption{$T$-dependence of the susceptibilities $\chi_{zz}\!=\!\lim_{H_z\to 0}\frac{m_z}{H_z}$ (a) and $\chi_{xx}\!=\!\lim_{H_x\to 0}\frac{m_x}{H_x}$ (b), for a spin-1/2 triangle described by $\mc{H}_{J}\!+\!\mc{H}_{\text{DM},z}+{\bf h}\cdot {\bf S}$, where ${\bf h}\!=\!g\mu_B{\bf H}$. The curves shown correspond to $|c|/J=0$ (black), 0.1 (blue) and 0.2 (red).}\label{fig:DMc-chi-vs-H}
\end{figure}

\section{Estimation of the DM parameters by fitting the experimental $\chi(T)$}\label{sec:ChiFit}
Here, we search for the set of parameters -- the Heisenberg exchange $J$ and
the DM exchanges $D_{\parallel}$ and $D_{\perp}$ -- of the isolated triangle
model that provide an optimal description of the experimental magnetic
susceptibility for both field directions. To this end, we use two complementary
approaches. In Sec.~\ref{sec:ChiFitTriangle}, we apply the analytical
expressions from Eqs.~(\ref{eq:chizz}) and (\ref{eq:chixx}), which are
valid for $D_{\parallel}=0$. In Sec.~\ref{sec:ChiFitED}, we calculate the
reduced magnetic susceptibilities numerically for different values of
$|\vec{D}|/J$ by keeping the $D_{\perp}/D_{\parallel}$ ratio fixed to
$\frac{\sqrt{3}}{3}$ $[$Eq.~(\ref{eq:dparallel_dperp})$]$. In both cases, the
minimal fitting temperature $T_{\text{min}}$ is used as a free parameter.

\subsection{Analytical expressions}\label{sec:ChiFitTriangle}
Analytical expressions in Eqs.~(\ref{eq:chizz}) and (\ref{eq:chixx}),
supplemented with a temperature-independent contribution $\chi_0$, account for
the experimental susceptibility down to approximately 15\,K
(Fig.~\ref{fig:chi_fit_triang}); at lower temperatures the discrepancies become
manifest (Fig.~\ref{fig:chi_dev_triang}, right). In the broad $T_{\text{min}}$
range between 12 and 20\,K, the fitted $J$ and $D_{\parallel}$  are nearly constant
(Fig.~\ref{fig:chi_dev_triang}, left and middle) and amount to 60\,K and 22 K,
respectively ($D_{\parallel}/J$ = 0.365).

\begin{figure}[h]
\centering
\includegraphics[width=.85\textwidth]{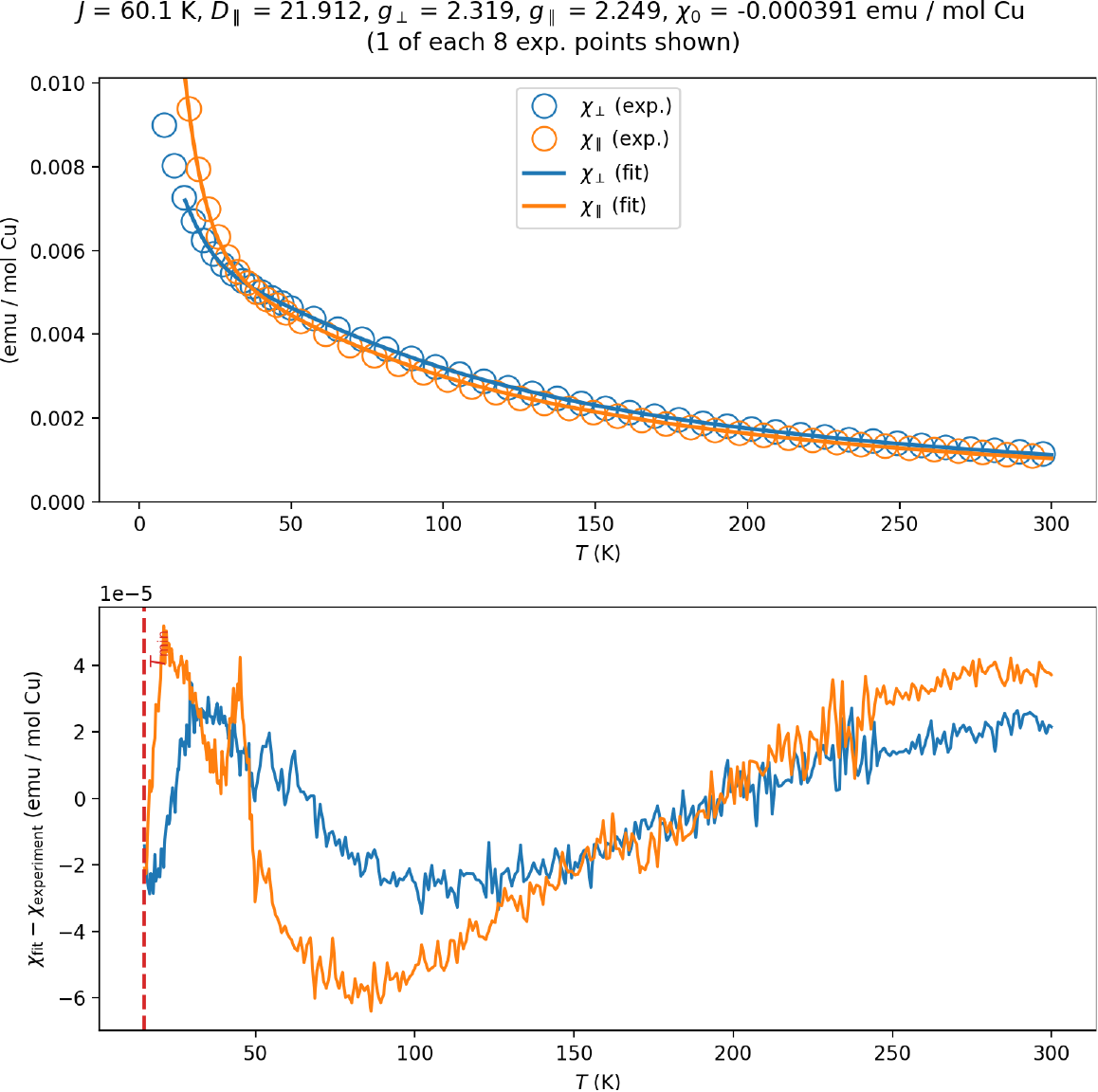}
\caption{\label{fig:chi_fit_triang}Top: Least-square fits to the experimental (circles) susceptibility (measured in 1\,T field) as function of temperature. Fitting is done for temperatures above $T_{\text{min}}$ = 15\,K. Bottom: difference plots for the same quantities. The plotting was made using matplotlib~\cite{sm:matplotlib}.
}
\end{figure}

\begin{figure}[h]
\centering
\includegraphics[width=.925\textwidth]{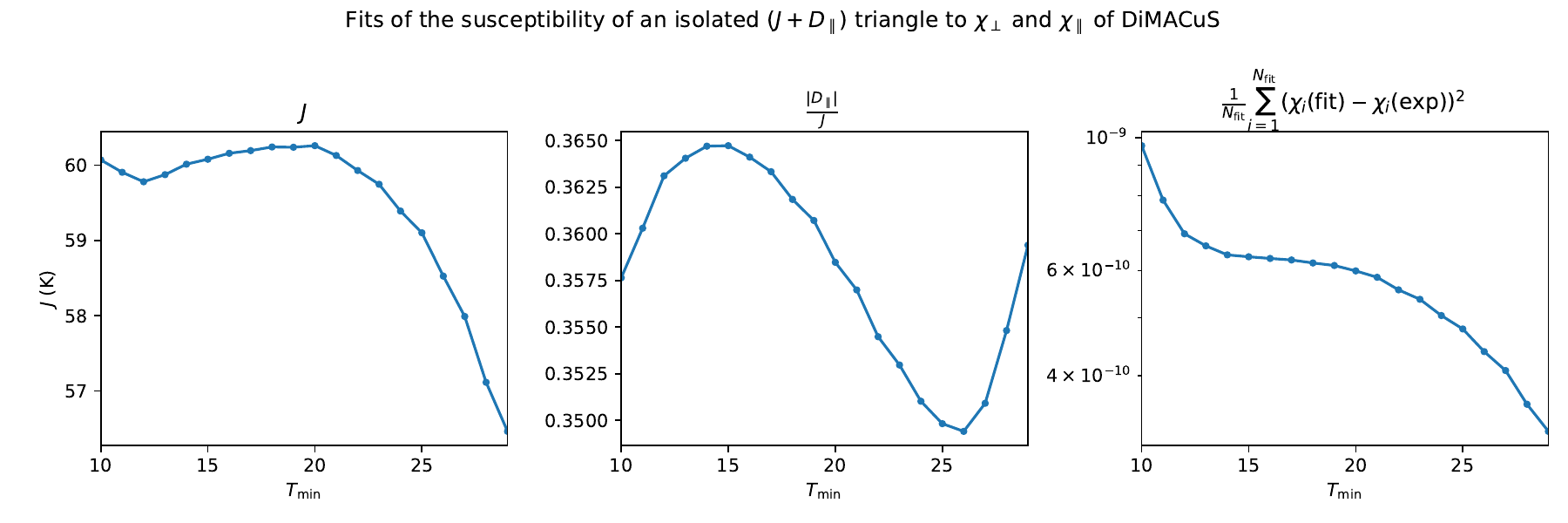}
\caption{\label{fig:chi_dev_triang}The fitted $J$ (left) and $D_{\parallel}/J$ (middle) as a function of the minimal fitting temperature $T_{\text{min}}$. Right: the normalized sum (over all fitted temperature and both field directions) of squared differences between the experimental and fitted $\chi(T)$ as a function of the minimal fitting temperature $T_{\text{min}}$. The plotting was made using matplotlib~\cite{sm:matplotlib}.}
\end{figure}

\subsection{Numerical evaluation}\label{sec:ChiFitED}
Here we include all the components of the DM vectors and calculate the magnetization and the magnetic susceptibility of a $S=\frac12$ spin triangle [Eq.~(\ref{eq:TriangleHam})] numerically. We work in energy units of $J=1$   and vary $|{\bf D}|/J$ between 0.2 an 0.6, while keeping the ratio $D_{\perp}/D_{\parallel}$ fixed to $\frac{\sqrt{3}}{3}$. The magnetic susceptibility is obtained by choosing the direction $\alpha$ ($x$ or $z$) of the reduced magnetic field $h^*=g\mu_BH/J$, calculating the reduced magnetization $M^*_{\alpha}$ and taking its zero-field derivative w.r.t.\
$h^*_{\alpha}$:
\be
M^*_{\alpha}(t^*, h^*) = -\frac{\operatorname{Tr}\left(S^{\alpha} e^{-\mc{H}(h^*)/t^*}\right)}{\operatorname{Tr}\left(e^{-\mc{H}(h^*)/t^*}\right)}\,,~~~
\chi^*_{\alpha\alpha}(t^*) = \Big(
\frac{\partial}{\partial h_{\alpha}^*} M^*_{\alpha}(t^*, h^*)
\Big)_
{h_\alpha^*\to0},
\ee
where $S^{\alpha}$ is the operator for the $\alpha$-th component of the spin, $M_\alpha^*=\frac{M_\alpha}{g\mu_B}$, $\chi_{\alpha\alpha}^*=\frac{\chi_{\alpha\alpha}}{(g\mu_B)^2/J}$,   and the reduced temperature $t^*=\frac{k_BT}{J}$ is taken on a mesh of 299 points between 0.04 and 6. The reduced susceptibilities $\chi^{*}_{xx}(t^*)$ and $\chi^{*}_{zz}(t^*)$ calculated on this $t^*$ mesh are approximated by a polynomial $\sum_i{c_{\alpha, i}(1/t^*)^i}$, where the coefficients $c_{\alpha, i}$ are determined numerically. The resulting polynomials are continuous functions of $t^*$ and can be fitted to the experiment using the following expression for the susceptibility of $N_A$ non-interacting spin triangles:
\be
\chi_{\alpha\alpha}(T) = \chi_0 + \frac{N_{\text{A}}(g \mu_{\text{B}})^2}{J}\sum_{i}{c_{\alpha, i} (1/t^*)^i},
\ee
where $J$, $g$, and $\chi_0$ are the fitting parameters that are
determined independently for $\alpha=x$ and $\alpha=z$.  In this way, we find
that $\chi_{\parallel}$ is largely independent of $|\vec{D}|/J$
(Fig.~\ref{fig:chi_dev_ed}, right). In contrast, $\chi_{\perp}$ is
sensitive to $|\vec{D}|/J$ (Fig.~\ref{fig:chi_dev_ed}, left), with the optimal
$|\vec{D}|/J\simeq$ 0.42, and hence $D_{\parallel}/J=0.364$ and
$D_{\perp}/J=0.21$.  These values are $25$\,\% smaller than the estimates from
first-principles calculations (Sec.~\ref{sec:j}). At the same time,
$D_{\parallel}/J$ is in remarkable agreement with the value obtained in the
analytical fits (Sec.~\ref{sec:ChiFitTriangle}), showing once again, that
the effect of the in-plane components of the DM vectors is weak, see
Sec.~\ref{sec:triangle}.

\begin{figure}[h]
\centering
\includegraphics[width=.675\textwidth]{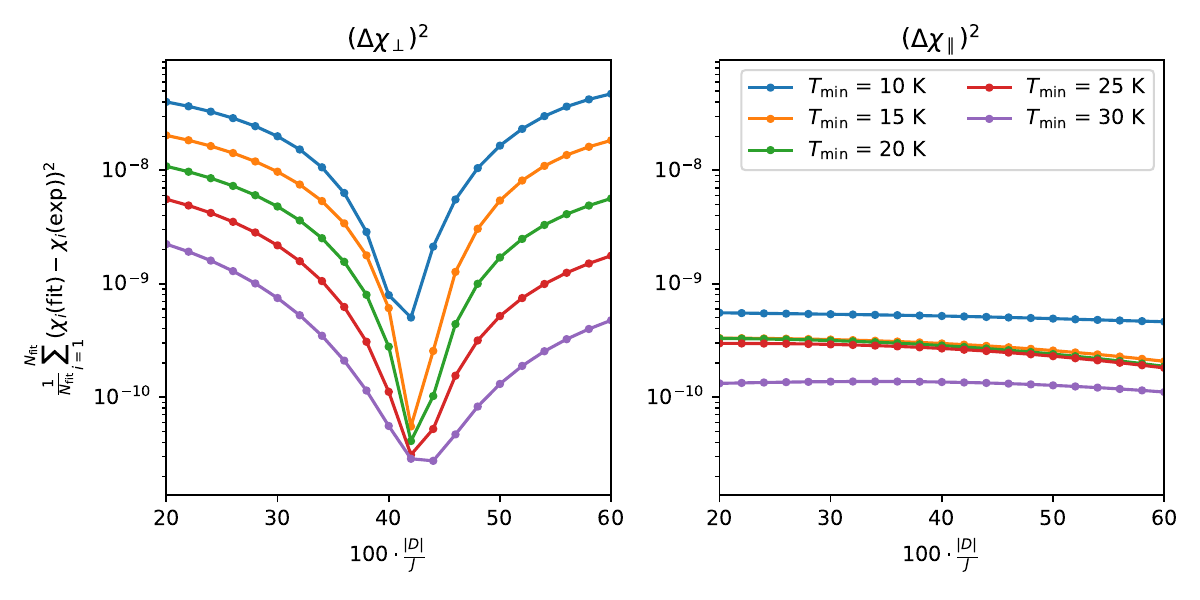}
\caption{\label{fig:chi_dev_ed}Normalized squared differences between the
calculated and experimental (measured 1\,T) magnetic susceptibilities
$\chi_{\perp}$ (left) and $\chi_{\parallel}$ (right) as a function of
$|\vec{D}|/J$ for various minimal fitting temperatures $T_{\text{min}}$. The plotting was made using matplotlib~\cite{sm:matplotlib}.}
\end{figure}

\subsection{Comparison with the isotropic model}\label{sec:ChiFitCompar}
Finally, we compare the optimal solution of the anisotropic model with the
isotropic Heisenberg model. While the out-of-plane susceptibilities can be
fitted by either model equally well (Fig.~\ref{fig:chi_iso_aniso}, middle
panels), only the anisotropic model can account for the in-plane susceptibility
(Fig.~\ref{fig:chi_iso_aniso}, top panels), which becomes evident if we inspect
the absolute difference (Fig.~\ref{fig:chi_iso_aniso}, bottom panels).

\begin{figure}[h]
\centering
\includegraphics[width=.85\textwidth]{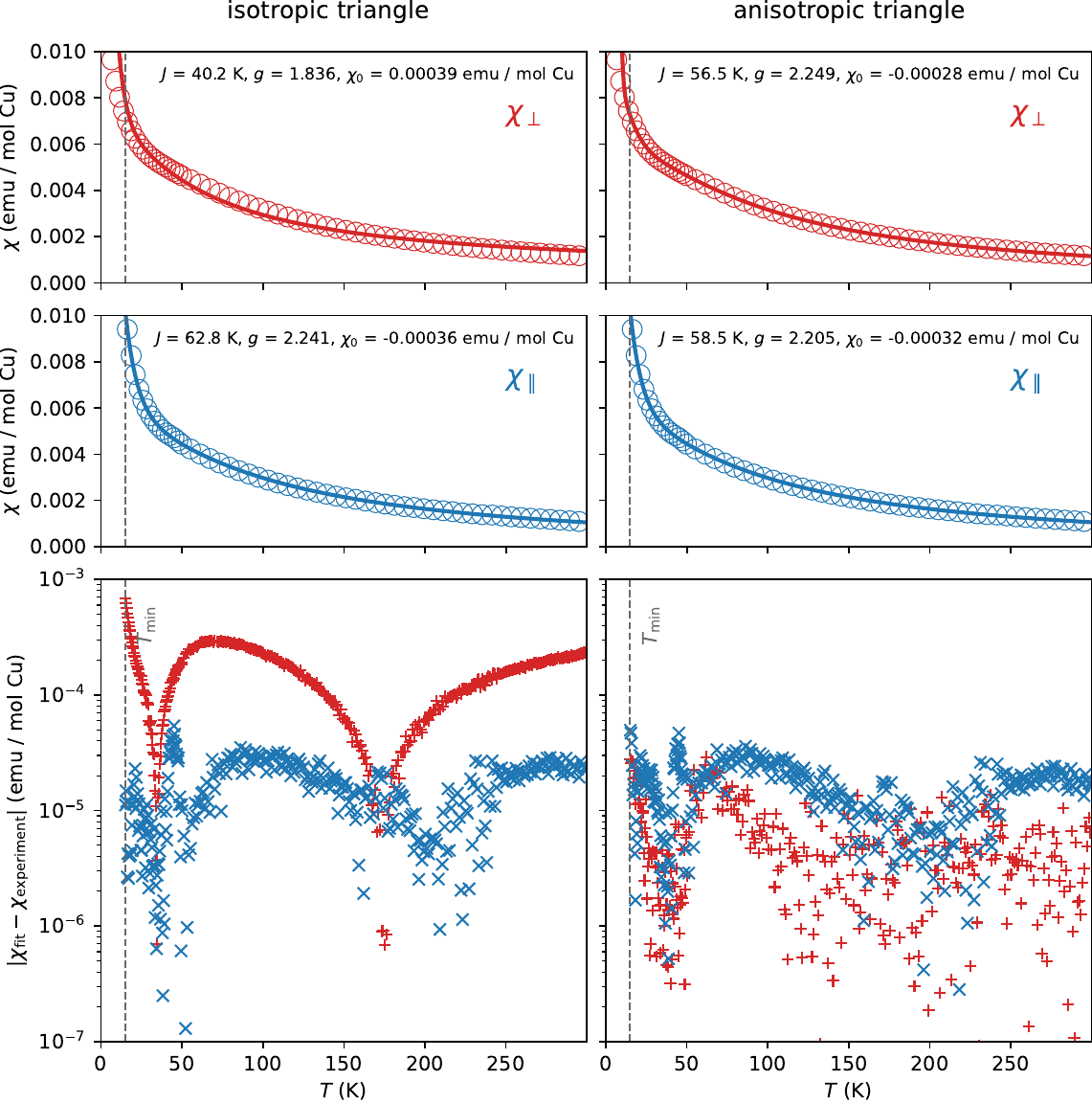}
\caption{\label{fig:chi_iso_aniso}
Fits to the experimental magnetic susceptibility with the isolated Heisenberg
triangle model (left) in comparison with the $|\vec{D}|/J=0.42,
D_{\perp}/D_{\parallel}=\frac{\sqrt{3}}{3}$ solution (Sec.~\ref{sec:ChiFitED})
of the anisotropic model (right). Top and middle panels show in-plane and
out-of-plane susceptibilities, bottom panels show the absolute difference
between the fit and the experiment.  The minimal fitting temperature
$T_{\text{min}}=15$\,K is denoted with a dashed line.  The plotting was made
using matplotlib~\cite{sm:matplotlib}.}
\end{figure}


%

\end{widetext}
\end{document}